%% file: mainv2.tex
\documentclass[10pt,a4paper,usenames,dvipsnames]{article}
\RequirePackage{pdf14}
\usepackage{jheppub}
\usepackage{comment} 

\usepackage{tikz}
\usetikzlibrary{shapes.misc}
\usetikzlibrary{arrows,automata}

\usepackage{makecell}
\usepackage{adjustbox}

\usepackage[utf8]{inputenc}
\usepackage{graphicx,tikz}
\usepackage{amsfonts,amsmath,amssymb}
\usepackage{array,setspace,mathrsfs,yfonts,dsfont,bbm,colonequals,amscd,mathtools}
\usepackage{relsize,suffix,cancel,capt-of,upgreek,verbatim,xcolor}
\usetikzlibrary{calc,shadings,patterns,tikzmark}
\usepackage{multirow,colortbl,rotating,arydshln} 
\usepackage{adjustbox}

\makeatletter
\gdef\@fpheader{\break}
\makeatother

\usepackage{amsthm}
\newtheorem{conjecture}{Conjecture}

\newtheorem*{definition-nonum}{Definition}

\newcommand{\ioni}{\mathfrak{n}}

\newcommand{\ioniUV}{\mathfrak{n}_{\text{UV}}}
\newcommand{\ioniIR}{\mathfrak{n}_{\text{IR}}}

\newcommand{\TMac}{\mathsf{T}}
\newcommand{\redTMac}{\textcolor{red}{\TMac}}
\newcommand{\n}{\mathfrak{n}}
\newcommand{\rank}{\mathfrak{rank}}

\title{The Nilpotency Index for 4d $\mathcal{N}=2$ SCFTs }

\author[a]{Anirudh Deb,\!}
\author[b]{Carlo Meneghelli,\!}
\author[a]{Leonardo Rastelli}

\affiliation[a]{C. N. Yang Institute for Theoretical Physics, Stony Brook University, Stony Brook, NY 11794-3840, USA}
\affiliation[b]{Dipartimento SMFI, Universit\`a di Parma, Viale G.P. Usberti 7/A, 43100, Parma, Italy and INFN Gruppo Collegato di Parma}

\emailAdd{anirudh.deb@stonybrook.edu}
\emailAdd{carlo.meneghelli@unipr.it}
\emailAdd{leonardo.rastelli@stonybrook.edu}

\begin{document}

\begin{flushright}
YITP-SB-2025-05
\end{flushright}

\abstract{
A well-developed classification program for 
4d $\mathcal{N}=2$ super conformal field theories (SCFTs) leverages  Seiberg-Witten geometry
on the Coulomb branch of vacua; theories are arranged by increasing $\mathfrak{rank}$,  the complex dimension of their Coulomb branch. 
An alternative  organizational scheme focusses on the associated vertex operator algebra (VOA), which is more closely related to the Higgs branch. 
From the VOA perspective, a natural way to arrange theories is by their ``index of nilpotency'', the smallest integer $\mathfrak{n}$ such that $T^\mathfrak{n} = 0$ in the
$C_2$ algebra, where $T$ is the VOA stress tensor.
It follows from the Higgs branch reconstruction conjecture that $\mathfrak{n} < \infty$ for any 4d ${\cal N}=2$ SCFT.
Extrapolating from several examples, we conjecture that  $\mathfrak{n}$ is an RG monotone, $\mathfrak{n}_{\rm IR} \leq \mathfrak{n}_{\rm UV}$. What's more, 
we find in all cases  that $\mathfrak{rank} \leq \mathfrak{n}-1$. Theory ordering by $\mathfrak{n}$ appears thus more refined than  ordering by $\mathfrak{rank}$. For example, in the list of $\mathfrak{rank}=1$
theories, the Kodaira SCFTs and $SU(2)$ ${\cal N}=4$ SYM  have $\mathfrak{n} =2$,
while all others have $\mathfrak{n} >2$.
}

\maketitle

\input{sections/S1_intro}

\input{sections/S2}

\input{sections/S3}
\input{sections/Sresults}

\input{sections/S_complexity}

\input{sections/S_RG}

\acknowledgments

 We would like to thank Tomoyuki Arakawa, Christopher Beem, Federico Bonetti, Thomas Creutzig, Jethro van Ekeren, Mario Martone and Gonenc Mogol  for helpful discussion. The work of AD and LR is supported in part by NSF grant PHY-2210533 and by the Simons Foundation grant 681267 (Simons Investigator Award).
The work of CM is supported in part by the Italian Ministero dell’Universit\`a e della Ricerca (MIUR), and by Istituto Nazionale di Fisica Nucleare (INFN)
through the ``Gauge and String Theory'' (GAST) research project.

\bibliographystyle{JHEP}
\bibliography{bibliref.bib}
\end{document}

%% file: sections/S1_intro.tex
\section{Introduction and summary}
\label{sec:introduction}
 Four-dimensional $\mathcal{N}=2$ superconformal field theories (SCFTs) display rich mathematical structures. A longstanding research program aims to leverage 
 the rigidity of these  structures to achieve a complete classification
 of 4d $\mathcal{N}=2$ SCFTs.
 There are in fact two, seemingly distinct, categories of mathematical invariants that are naturally associated to 4d  $\mathcal{N}=2$ SCFTs.
 One is the special K\"ahler geometry~\cite{Seiberg_1994,Seiberg_1994(2)} of their Coulomb branch of vacua (see \cite{Martone:2020hvy,Argyres:2024uuc} for recent overviews). The other is the vertex operator algebra (VOA) that can be carved out from the full local operator algebra by a cohomological reduction~\cite{Beem_2015}. As a vector space, the VOA comprises the space of Schur operators, obeying a certain shortening condition for the ${\cal N}=2$ superconformal algebra. 
 The VOA can be viewed as a vast (non-commutative) extension  of the Higgs branch chiral ring -- indeed Higgs chiral operators (whose vevs parametrize the Higgs branch of vacua) are a special type of Schur operators. A central conjecture~\cite{Beem_2018} is that the Higgs branch (as a  holomorphic symplectic variety)  can be recovered from the VOA by the canonical mathematical procedure of computing its ``associated variety''~\cite{arakawa2012remark}.

\begin{figure}[t]
 \centering
\begin{tikzpicture}[scale=0.9,
->,
  >=stealth',
  shorten >=1pt,
  auto,
  node distance=2.8cm,
  semithick]
    \node[] at (0, 0)   (a) {4d $\mathcal{N}=2$ SCFT};
    \node[] at (-4.5,-4)   (b) {\shortstack{Seiberg-Witten Geometry \\``Coulomb Branch"}};
    \node[] at (4.5,-4)   (c) {\shortstack{Vertex Operator Algebra \\``Higgs Branch"}};
    \node[] at (0,-8)   (e) {Schur Index};

    \path[every node/.style={sloped,anchor=south,auto=false}]
        (a) edge [draw=violet]             node {} (b)         
        (a) edge[draw=violet]               node {} (c)
        (b) edge[draw=violet]               node {\large\textcolor{red}{?}} (c)
        (c) edge[draw=violet]               node {} (b)
        (b) edge[draw=violet]            node {\footnotesize \textcolor{blue}{Wall Crossing Invariant}} (e)
        (c) edge[draw=violet]               node {\footnotesize \textcolor{blue}{Vacuum Character}} (e);
\end{tikzpicture}
\caption{The diagram depicts some of the connections  between the Coulomb and the Higgs branches and their associated mathematical structures.
}
\label{fig:relationCBHBVOA}
\end{figure}
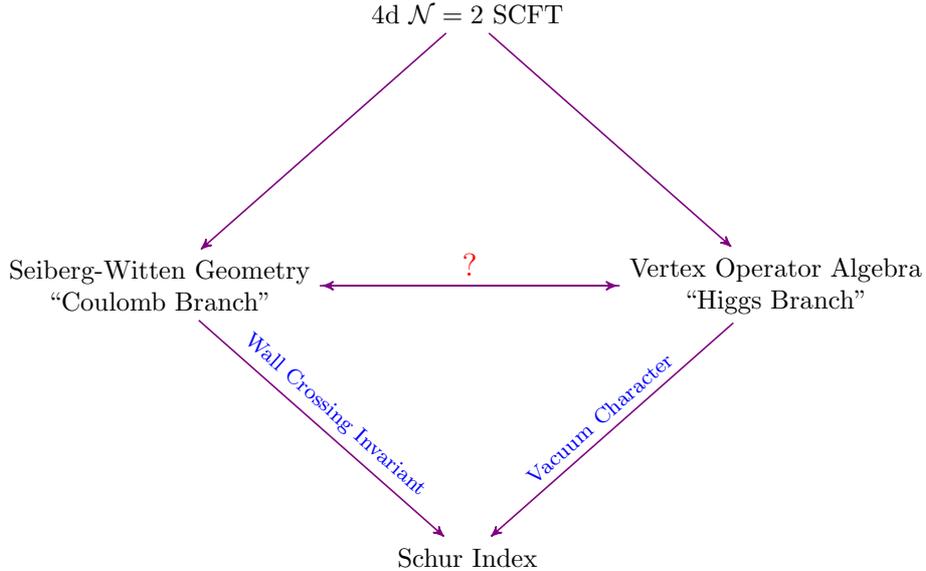

  Each of these two invariants
  appears to be separately sufficient to fully characterize the 4d ${\cal N}=2$ SCFT, indeed to the best of  our knowledge no two distinct\footnote{A disclaimer is in order. We are focussing here on the local operator algebra on $\mathbb{R}^4$: two theories with the same spectrum of local operators and the same OPE are counted as the same theory. We are  ignoring the finer distinctions that include the spectrum of extended operators or global properties in nontrivial geometries. For Lagrangian theories, these are the distinctions associated to the global choice of  Lie group of gauge transformations.} 4d ${\cal N}=2$ SCFTS have the same Coulomb branch (CB) geometry or the same VOA. Given a VOA that descends from a 4d theory, we should be able to construct a map to the CB geometry, and vice versa. 
This is surprising, as the two structures belong to very different mathematical realms.
A remarkable hint of a correspondence between the two classes of invariants comes from the observation~\cite{VPA3} (inspired by~\cite{Cecotti:2010fi}) 
that the Schur index, which is directly identified with the vacuum character of the VOA, can also be computed in terms of a wall-crossing invariant trace that counts massive BPS particles on the Coulomb branch. See Figure \ref{fig:relationCBHBVOA}. Other subtle examples of mutual compatibility between the two structures were encountered in~\cite{Kaidi:2022sng, Buican:2024jjl,Fredrickson:2017yka, Pan:2024hcz}.

  In CB geometry, it is natural to 
order theories according to
the complex dimension of their CB,  known as the $\rank$. (The name stems from the fact that in a Lagrangian theory, 
this is just the rank of the gauge group.)
A complete classification of consistent CB geometries has been achieved~\cite{Argyres:2015ffa, Argyres:2015gha,Argyres:2016xmc,Argyres:2016xua,Argyres:2020nrr} for $\mathfrak{rank}=1$;
they have been  identified with known SCFTs, up to a couple of exotic exceptions.
A comprehensive analysis (close to complete) has by now also been carried out~\cite{Argyres:2022lah,Argyres:2022puv,Argyres:2022fwy} for $\mathfrak{rank}=2$. From the VOA perspective, it is at first less obvious how to organize theories by increasing degree of complexity. A possible approach 
was suggested in \cite{Beem_2018}. A
consequence of the Higgs branch reconstruction conjecture is that the vacuum module of the VOA (the Schur index) must obey a monic modular differential equation (MDE). Organizing theories by the {\it order} $\mathfrak{ord}(\mathcal{D})$ of the modular differential operator seems a promising route. Indeed this is the approach taken in \cite{Kaidi:2022sng}, which carried out a systematic scan of possible solutions of low-order MDEs with integral coefficients, subject to a variety of other consistency conditions (including some coming from the CB). Here we propose to organize theories by a conceptually related but more intrinsic quantity,
the ``nilpotency index'' $\mathfrak{n}$ of the VOA stress tensor.\footnote{
We warn our readers that the moniker ``nilpotency index'' has appeared before~\cite{Komargodski:2020ved} in the superconformal field theory literature, to denote an invariant of conformal manifolds in theories with four supercharges; needless to say, there is no relation between their use and ours. 
}

The statement that the VOA stress tensor does  {\it not} correspond to a Higgs chiral operator translates~\cite{Beem_2018} into the requirement that $T^\mathfrak{n}$, for some positive integer $\mathfrak{n}$ (defined as the smallest such integer), must be expressible in terms of normal ordered products that contain at least one holomorphic derivative. In mathematical parlance, ``$T$ is nilpotent in the $C_2$ algebra'', with nilpotency index $\mathfrak{n}$. As we review below,
it is expected that $\mathfrak{n} \leq \mathfrak{ord}({\cal D})$.

Our investigation was sparked by the striking observation that in several simple examples,
$\mathfrak{rank} = \mathfrak{n}-1$, a really  unexpected relation between integer invariants that are superficially completely unrelated! We set out to consider more elaborate examples, and found that the above relation can be violated, but seemingly always in the same direction.  We are led to the following conjecture:
  \begin{conjecture} 
  	\label{conj:1}
  	The nilpotency  index $\mathfrak{n}$ and the
    $\mathfrak{rank}$ of a 4d ${\cal N}=2$ SCFT obey the inequality
   \begin{equation}
    \mathfrak{rank} \leq \mathfrak{n}-1
\end{equation}
 \end{conjecture}
If  this conjecture is true, it implies that theory ordering by $\mathfrak{n}$ is more refined than ordering by $\mathfrak{rank}$.  We show  that the only SCFTs with $\mathfrak{n}=1$ are collections of free hypermultiplets, which are  believed (general lore!) to be the only rank-zero SCFTs, so that the inequality is saturated. Things are more interesting for rank one. We find
that the Kodaira SCFTs (i.e.\ the Deligne series of Section \ref{sec:drankone}) and $SU(2)$ ${\cal N}=4$ SYM  have $\mathfrak{n} =2$,
while all others have $\mathfrak{n} >2$.

 In some sense, the nilpotency index (much like the rank, but in a more refined way if  conjecture~1 is true) is a measure of complexity of the SCFT. It is natural to ask whether this can be made sharper: is $\mathfrak{n}$ an RG monotone? We have tested this idea in a few (admittedly not very many) examples of RG flows. We have considered both flows triggered by relevant deformations and flows triggered by moving on the moduli space, and found that our hypothesis indeed holds. We thus propose:
  \begin{conjecture}
  \label{conj:2}
  If two 4d $\mathcal{N}=2$ SCFTs are related by RG flow, then the nilpotency index of the IR theory is always smaller than or equal to the nilpotency index of the UV theory, 
  \begin{equation}
    \mathfrak{n}_{\rm IR} \leq \mathfrak{n}_{\rm UV} \,.
\end{equation}
 \end{conjecture}
 In fact conjecture 2 implies conjecture 1. 
This follows simply by looking at the infrared theory at a generic point on the Coulomb branch,
which consists of
just $\text{rank}$ free vector multiplets,
for which it is elementary to show 
that $\n_{\rm IR} =\mathfrak{rank} +1$.

 A special case of conjecture 2 is nilpotent Higgsing of a  4d ${\cal N}=2 $ SCFT. For a theory with flavor algebra $\mathfrak{g}$, nilpotent higgsing is implemented by giving  a vev to the component of the moment map operator corresponding to a nilpotent element $e\in \mathfrak{g}$, the image of the raising generator of some embedding $\mathfrak{su}(2)\hookrightarrow\mathfrak{g}$. From the VOA perspective, this corresponds to the procedure of Drinfeld-Sokolov (DS) reduction \cite{Drinfeld:1984qv,deBoer:1992sy,deBoer:1993iz,Feigin:1990pn,Feher:1992ed,Beem:2014rza}. It seems worthwhile to specialize conjecture~2 to this case, as it becomes a statement that can be phrased purely in VOA language:
\begin{conjecture}
\label{conj:3}
The nilpotency index $\n$ is non-increasing under Drinfeld-Sokolov reduction.
\end{conjecture}
While we have only checked this conjecture in examples of VOAs that arise from 4d $\mathcal{N}=2$ SCFTs, one may wonder whether it holds more generally for any VOA.
In fact, we give a sketch of a scheme theoretic proof in Section \ref{sec:schemeproof}.

The concrete results that support our conjectures are summarized in table~\ref{tab:summaryres}. (See also table~\ref{tab:ionirankone} for some additional  information about rank-one theories.) Perhaps the most interesting question raised by our findings is {\it why}
any of this should be true. We are seeing yet another hint of a connection between the VOA and the CB structures but a deeper explanation remains elusive. In terms of concrete directions, there is the natural program of classifying VOAs 
(associated to 4d SCFTs)  ordering them
by increasing values of $\n$. Here we illustrate this idea in  the simple cases of $\n =1$ and $\n=2$. For $\n =2$ we have achieved a classification under a natural hypothesis about the $R$-filtration of certain composite operators in the VOA, but it should be possible to relax this assumption by a more systematic analysis. It would then be very interesting to tackle the classification problem for  $\n=3$. Several rank-one theories and some special rank-two theories are expected to show up.
We would be particularly curious to know the list of rank-two theories with $\n=3$ (the minimal allowed value, if our conjecture~1 is correct), and to understand what makes them special.

The paper is organized as follows. In Section~\ref{sec:IndexofNilpotency} we review the very basics of  the VOA/SCFT correspondence and then define the nilpotency index. We  illustrate it in the elementary examples of the free hyper and the free vector multiplets.
Section \ref{sec:Methods} explains different methods to determine the nilpotency index. The most straightforward is to find the requisite null by computing norms, but this is often hard, in which case alternative tools  are useful. 
In Section~\ref{sec:resultsdiscuss} we  discuss the various examples of $\mathcal{N}=2$ SCFTs for which we have determined the nilpotency index.  In Section \ref{sec:measureofcomplexity} we classify the theories with $\ioni=1$ and (under a certain natural assumption) the ones with  $\ioni=2$. In Section \ref{sec:RGFlow} we discuss  our experimental observation that the nilpotency index appears to be non-increasing under RG flow. 

\begin{table}[t]
	\begin{adjustbox}{center}
		\begin{tabular}{ |c|c|c|c|c|c| }
                \hline
				Theory & $T^\mathfrak{n}$ & Rank & MLDE & $c_{2d}=-12c_{4d}$ & $a_{4d}$ \\ 
				
				\hline
                \hline
				$SU(2)$ SQCD & $T^2$  &$1$  & $2$ U& $-14$ & $\frac{23}{24}$  \\
				\hline
				$SU(2)$  $\mathcal{N}=4 $ SYM & $T^2$  &$1$  & $2$ T& $-9$ & $\frac{3}{4}$  \\		
                \hline
				Deligne rank-one  & $T^{2}$ & $1$ & $2$  U  & $-2-2h^\vee$ & $\frac{5+3h^\vee}{24}$ \\
				\hline
				\hline
			
				$(A_1,A_{2n})$ & $T^{n+1}$ & $n$ & $n+1$  U & $-\frac{2n(6n+5)}{2n+3}$ & $\frac{n(24n+19)}{24(2n+3)}$ \\ 
				\hline
				\hline
                \rowcolor{orange!20}
				$(A_1,D_{2n+1})$ & $T^{n+1}$ & $n$ & $n+1$ &  $-6n$ & $\frac{n(8n+3)}{8(2n+1)}$ \\
                \hline
				$(A_1,D_3)\sim(A_1,A_3)$ & $T^{2}$ & $1$ & $2$ U    & $-6$ &  $\frac{11}{24}$ \\
                \hline
                $(A_1,D_5)$ & $T^{3}$ & $2$ & $3$ U  & $-12$ &  $\frac{19}{20}$ \\
				\hline
				\hline
                \rowcolor{orange!20}
				$(A_1,A_{2n+1})$ & $T^{n+1}$ & $n$ &   & $-\frac{2 \left(3 n^2+5 n+1\right)}{n+2}$ &  $\frac{12 n^2+19 n+2}{24 n+48}$ \\
				\hline
				$(A_1,A_5)$ & $T^{3}$ & $2$ & $4$ T    & $-\frac{23}{2}$ &  $\frac{11}{12}$ \\
				\hline
				
				$(A_1,A_7)$ & $T^4$  & $3$ & $6$ U & $-\frac{86}{5}$ &  $\frac{167}{120}$\\
				\hline
				\hline
		    \rowcolor{orange!20}
				$(A_1,D_{2n+2})$ & $T^{n+1}$ & $n$ &   & $-6 n-2$ & $\frac{6n+1}{12}$ \\
                \hline
				$(A_1,D_{4})$& $T^2$ & 1 & 2  U  & $-8$ & $\frac{7}{12}$ \\
				\hline
				
				$(A_1,D_{6})$ & $T^3$ & 2 & 4  T  &$-14$ & $\frac{13}{12}$ \\
				
				\hline
				\hline
    
				$(A_1,E_6)\sim (A_2,A_3)$ & $T^{4}$ & $3$ & $4$  U  & $-\frac{114}{7}$ & $\frac{75}{56}$ \\
				\hline
				
				$(A_1,E_8)\sim (A_2,A_4)$ & $T^5$ & $4$ & $5$  U  & $-23$ & $\frac{91}{48}$ \\
				\hline
				
                \hline
				$\mathcal{N}=3$ $\mathbb{C}^3/\mathbb{Z}_3$ & $T^{3}$  &$1$ & $4$ T& $-15$ & $\frac{5}{4}$ \\
				$\mathcal{N}=3$ $\mathbb{C}^3/\mathbb{Z}_4$ & $T^{4}$  &$1$  & $5$ T& $-21$ & $\frac{7}{4}$ \\
				\hline
				\hline
                \rowcolor{yellow!20}
				$SU(3)$ SQCD & $T^{3}$ & $2$  & $4$ T & $-34$ & $\frac{29}{12}$ \\
				\hline
                \rowcolor{yellow!20}
                $SU(3)$  $\mathcal{N}=4 $ SYM & $T^3$ & $2$  & $4$ U & -24 & 2 \\ 
				\hline
                \rowcolor{yellow!20}
				Deligne Rank 2  $\mathfrak{d}_4$ & $T^{3}$ & $2$ & $4$ T &$-41$ & $\frac{71}{24}$ \\
				\hline
                \rowcolor{yellow!20}
				Class S $A_1$ quivers & $T^{s-2}$ & $s-3$ & \tiny see \cite{Beem_2022} \normalsize T/U & $-2(5s-13)$ & $\frac{19s-53}{24}$ \\
				\hline
		\end{tabular}
		
	\end{adjustbox}

	\caption{This table summarizes the data of the theories discussed in Section \ref{sec:resultsdiscuss}. The second column displays the nilpotency index $\ioni$. The MLDE column states the order of MLDE and whether it is a twisted (T) or untwisted (U) MLDE. The rows in orange are theories which come in families labelled by an integer ``$n$"  and the value of $\ioni$ is conjectural based on the examples checked, which are given in white rows immediately below them. For such cases finding the null gets difficult as ``$n$" increases and the conjectural value for higher ``$n$" is based on Macdonald indices. In certain places the MLDE data are left blank to indicate we don't have formula for the order of the MLDE in terms of the parameter labelling the family of theories. The rows in yellow are for theories where we haven't found an explicit null, but we can conjecture  $\ioni$ from the Macdonald index. The order of the MLDE for class S $A_1$ quivers is discussed in \cite{Beem_2022}.}
	\label{tab:summaryres}
\end{table}

%% file: sections/S2.tex
\section{Nilpotency index}
\label{sec:IndexofNilpotency}

In this section, after a lightning review of the VOA/SCFT correspondence, we define the nilpotency index. We  then illustrate the definition in the elementary examples of the free vector multiplet and free hypermultiplets. Finally we discuss some general structural features of the nilpotency index for tensor products and discrete gaugings of theories.

\subsection{VOA/SCFT correspondence and Higgs branch conjecture}

 A 4d $\mathcal{N}=2$ SCFT has an associated vertex operator algebra \cite{Beem_2015}
    \begin{equation*}
        \text{4d SCFT}\xrightarrow[]{\mathbb{V}}\text{VOA}\;,
    \end{equation*}
obtained by passing to the cohomology of a certain nilpotent supercharge. As a vector space, the VOA consists of the Schur operators whose quantum numbers satisfy
\begin{equation}
\label{eq:schurcondition}
\begin{split}
    E-(j_1+j_2)-2R&=0\;,\\
    r+j_1-j_2&=0\;,
\end{split}
\end{equation}
where $E$ is the conformal dimension, 
$j_1$ and $j_2$  the Cartan eigenvalues of the $SU(2)\times SU(2)$ rotations,  
$R$ and $r$ the Cartan eigenvalues of the $SU(2)_R$  and  $U(1)_r$ R-symmetries.  These  are the operators that contribute to the Schur  limit of the superconformal index~\cite{Gadde:2011ik,Gadde:2011uv, Rastelli:2014jja}. 

 Every local 4d SCFT has a stress tensor multiplet which contains the $SU(2)_R$ currents. The leading pole of the self OPE of these currents is proportional to the Weyl anomaly $c_{4d}$. The VOA image of the $SU(2)_R$ current is the stress tensor, which obeys the standard two dimensional OPEs with central charge $c_{2d}$. The central charge of the 4d SCFT and the 2d VOA are related to each other in the following way
\begin{equation}
    c_{2d}=-12c_{4d}\;.
\end{equation}
If the SCFT also has a flavor symmetry, the operators surviving the cohomology are the moment maps. Similar to the stress energy tensor case, the leading pole of the self OPE of the moment maps is proportional to four-dimensional flavor central charge $k_{4d}$ and the VOA image of the moment maps has OPEs that of the affine Kac-Moody (AKM) algebra with central charge $k_{2d}$, which is related to $k_{4d}$ as follows
\begin{equation}
	 k_{2d}=-\frac{1}{2}k_{4d}\;.
\end{equation}
 
The holomorphic conformal dimension of the VOA operators is related to four-dimensional quantum numbers as follows
\begin{equation}
    h=\frac{E+j_1+j_2}{2}=E-R~.
\end{equation}
Using the expression for the conformal dimension and the relation between $c_{4d}$ and $c_{2d}$, one can show that the vacuum character of the VOA equals the Schur index of the SCFT up to an overall power of $q$. The vacuum character $\chi_{\text{vac}}(q)$ and the Schur index $\mathcal{I}_{\text{Schur}}(q)$ are related as follows
\begin{equation}
    \chi_{\text{vac}}(q)= \text{Tr}(-1)^F q^{h-\frac{c_{2d}}{24}}~,
\end{equation}
\begin{equation} \label{SchurVac}
    \mathcal{I}_{\text{Schur}}(q)=\text{Tr}(-1)^F q^{E-R}=\text{Tr}(-1)^F q^{h}= q^{-\frac{c_{4d}}{2}}\chi_{\text{vac}}(q)~.
\end{equation}
(See Section \ref{sec:superconformalindex} for more details on the superconformal index). 

Besides the holomorphic scaling dimension $h$,
the space of Schur operators is graded by two additional quantum numbers, which we can take to be the R-charges $r$ and $R$. However, while $r$ is respected by the VOA structure (it simply adds under normal ordered product), $R$ is not. Because
the cohomological construction involves lower components under  $SU(2)_R$, normal ordered products have a value of $R$ that is in general smaller than or equal to the sum of the values of their constituents. We will refer to this filtration as the ``$R$-filtration'' of the VOA. 
  
We will assume that our VOAs have finitely many \textit{strong generators}. Strong generators are operators which cannot be expressed as normal ordered products of other operators and their derivatives. We will denote the normal ordered product of $A$ and $B$ simply as $AB$. The $n$-fold normal ordered product of an operator $A$ is denoted by $A^n$.

Let us now review the Higgs branch reconstruction conjecture \cite{Beem_2018}. We regard the Higgs branch ${\cal M}_H$ as a holomorphic symplectic variety. Its ring of holomorphic functions $\mathbb{C}[{\cal M}_{H}]$ is a reduced, commutative, associative $\mathbb{C}$-algebra equipped with a Poisson bracket. By the standard QFT lore,  we identify $\mathbb{C}[{\cal M}_{H}]$ with the ring ${\cal R}_{H}$ of Higgs chiral operators. By definition, Higgs chiral operators obey $E =2 R$, 
$j_1 = j_2 = r =0$
and are of course a special class of Schur operators.
The Poisson bracket on ${\cal R}_H$ is defined in terms of a canonical secondary product. We have then
\begin{equation}
 \mathbb{C}[{\cal M}_H] = {\cal R}_H\,, 
\end{equation}
or equivalently (by the standard algebraic geometric dictionary)
\begin{equation}
{\cal M}_H = {\rm Specm} ({\cal R}_H) \,.
\end{equation}
We seek to reconstruct 
${\cal R}_H$ (equivalently ${\cal M}_H$)
from the VOA data. It turns out that to any VOA ${\cal V}$, 
one can associate~\cite{zhu1990vertex} a canonical commutative associative algebra ${\cal R}_{\cal V}$,
known as the Zhu's $C_2$-algebra or Zhu's commutative algebra.
Informally, ${\cal R}_{\cal V}$ is obtained from ${\cal V}$ by dropping
normal ordered composite operators containing at least one derivative. To wit, we first introduce $C_2(\mathcal{V})$,  the space of operators with at least one derivative, 
\begin{equation}
    C_2(\mathcal{V})\coloneqq\text{Span}\{u^i_{-h_i-1}\phi\,\,,\,\,u^i,\phi\in\mathcal{V}\}\;.
\end{equation}
 $\mathcal{R}_\mathcal{V}$
is then defined to be the quotient
\begin{equation}
\label{ZhuC2}
\mathcal{R}_\mathcal{V}\coloneqq\mathcal{V}/{C_2(\mathcal{V})}\;.
\end{equation}
One can show that $\mathcal{R}_{\mathcal{V}}$ is a commutative and associative algebra. It is  endowed
with a Poisson bracket that descends from the first order poles in the OPE. 
The $\mathcal{R}_{\mathcal{V}}$ algebra is in general {\it not} reduced.
The Higgs branch reconstruction conjecture~\cite{Beem_2018} simply
asserts that its reduced part (obtained by modding out by the nilradical) is the Higgs branch chiral ring,
\begin{equation}
{\cal R}_H = (\mathcal{R}_{\mathcal{V}})_{\rm red}\,.
\end{equation}
The equivalent geometric statement is
\begin{equation}
{\cal M}_{H} = {\rm Specm} ({\cal R}_H)= {\rm Specm} (\mathcal{R}_{\mathcal{V}})  \eqqcolon{\cal X}_{\cal V}\,.
\end{equation}
The variety ${\cal X}_{\cal V}$ was previously defined in \cite{arakawa2012remark} and is known as  the \textit{associated variety} of the VOA.
In summary, the Higgs branch reconstruction conjecture identifies the Higgs branch of the 4d SCFT with the associated variety of the corresponding VOA.

\subsection{Nilpotency index}

The nilpotency index is defined using a specific  null state containing a power of the stress tensor as one of its term. As the 4d stress tensor multiplet is not a Higgs branch multiplet,  the Higgs branch reconstruction conjecture implies that its cohomological image in the VOA should be nilpotent in $\mathcal{R}_\mathcal{V}$~\cite{Beem_2018}. We 
are ready for the following 
\begin{definition-nonum} 
The nilpotency index is
the smallest positive integer $\n$ for which the VOA stress tensor  $T^\n=0$ in
$\mathcal{R}_\mathcal{V}$, $
    T^\n \in C_2 ({\cal V})$.
    There is a corresponding null state $\mathfrak{N}$,
 \begin{equation}
 \label{nilpotencynull}
  \mathfrak{N}=  T^\n  + \phi \, , \quad \phi \in C_2 ({\cal V})\,
    \end{equation}
  which we call the ``nilpotency null''. 
 \end{definition-nonum}
\subsection{Nilpotency index  and modular linear differential equation}
The Schur index of the SCFT  is directly related (equation (\ref{SchurVac})) to the vacuum character of the corresponding VOA. 
Assuming the Higgs branch reconstruction conjecture, one can prove that the VOA vacuum character must obey a monic modular linear differential equation (MLDE)~\cite{arakawa2017quasilisse, Beem_2018}.
In favorable cases, the MLDE 
arises directly from the existence of the nilpotency null (\ref{nilpotencynull}):
a certain recursion relation\footnote{For a quick review see e.g.~section 3.1 and appendix of B of~\cite{Beem_2018}, and references therein.} allows to translate the vanishing of the torus one-point function  of $\mathfrak{N}$ into an order $\n$ modular differential operator ${\cal D}$ that annihilates the vacuum character. One may  however encounter a certain obstruction (as reviewed in section~3.1 of \cite{Beem_2018}) in carrying out one of the recursive steps. 
Nevertheless, if the Higgs branch reconstruction  conjecture is true, all VOAs that arise from 4d SCFTs are ``quasi-lisse'', which by definition means that their associated variety contains a finite number of symplectic leaves. (Indeed the Higgs branch is endowed with a nondegenerate symplectic form.)
An important result of~\cite{arakawa2017quasilisse} establishes the existence of a monic MLDE for the vacuum character for {\it all} quasi-lisse VOAs. Generically,\footnote{We say ``generically'' as we do not know of a rigorous argument that rules out the existence of an ``accidental'' modular differential operator completely unrelated to nilpotency of $T$.} one expects
\begin{equation}
\label{eq:inequlaityordnrank}
\n \leq \mathfrak{ord}(\mathcal{D}) \,.
\end{equation}
We have checked that this inequality holds  in all examples of VOAs associated to 4d SCFTs where both integers are known. 
\subsection{Elementary examples}
We now consider the two elementary examples
of free SCFTs.

\subsubsection{Free hypermultiplets}
The VOA of $k$ free hypermultiplets is just $k$ copies of $\beta\gamma$ system with both $\beta$ and $\gamma$ having weights $\frac{1}{2}$ and OPEs given as
\begin{equation}
    \beta_i(z)\gamma^j(w)\sim\frac{-\delta^j_i}{z-w},\;\;\;i,j=1, 2,\dots, k\;.
\end{equation}
The stress tensor of this VOA is not an independent generator and is given by 
\begin{equation}
    T=\frac{1}{2}(\partial\beta_i)\gamma^i-\frac{1}{2}\beta_i\partial\gamma^i\;.
\end{equation}
Since $\frac{1}{2}(\partial\beta_i)\gamma^i-\frac{1}{2}\beta_i\partial\gamma^i\in C_2(\mathcal{V})$, the above equation gives $\mathfrak{n}=1$,
\begin{equation}
    T^1=0\quad {\rm in}\; {\cal R}_{\cal V} \, .
\end{equation}
This is  of course consistent with conjecture~\ref{conj:1}, as the rank of free hypermultiplets is just zero. The vacuum character solves a first order differential equation, consistent with~\eqref{eq:inequlaityordnrank}.

\subsubsection{Free vector multiplets}
\label{sec:sympferm}
The VOA of $k$ free vector multiplets is just $k$ copies of symplectic fermions $\lambda^{i=1,2,\dots k}_{\alpha=1,2}$ of weight one and the following OPEs
\begin{equation}
		\lambda^i_\alpha(z)\lambda^j_\beta(w)\sim\frac{\epsilon_{\alpha\beta}\delta^{ij}}{(z-w)^2}\;.
\end{equation}
The stress tensor of this VOA is not an independent generator and is given by
\begin{equation}
    T=-\frac{1}{2}\epsilon^{\alpha\beta}\delta_{ij}\lambda^i_\alpha\lambda^j_\beta\;.
\end{equation}
Due to the fermionic nature of $\lambda_\alpha$, it is a simple computation to show that the $(k+1)$-fold  normal ordered product of the stress tensor lies in $C_2(\mathcal{V})$.
\begin{equation}
    T^{k+1}\in C_2(\mathcal{V})
\end{equation}
and therefore $\mathfrak{n}=k+1$, which is consistent with the conjectured inequality since rank of the $k$ free $U(1)$ vector multiplets is $k$. The vacuum character solves an MLDE of order $k+1$,
which is consistent with \eqref{eq:inequlaityordnrank}.

\subsection{Tensor products}
\label{sec:tenstheor}
We now discuss the nilpotency index for the tensor product of theories.
Consider first two decoupled theories $\mathcal{T}_1$ and $\mathcal{T}_2$, with individual stress tensors $T_1$ and $T_2$  having nilpotency indices $\mathfrak{n}_1$ and $\mathfrak{n}_2$ respectively. We have of course
\begin{equation}
\left(T_1+T_2\right)^{n}=\sum_{m=0}^n\binom{n}{m}T_1^m T_2^{n-m}\;.
\end{equation}
Clearly, if $m \geq  \mathfrak{n}_1$, 
or $n-m \geq \mathfrak{n}_2$, 
the monomial $T_1^m T_2^{n-m}$ is zero.
Consider the ``worst case'' for $T_1$, namely the monomial $T_1^{\mathfrak{n}_1-1} T_2^{n - \mathfrak{n}_1 + 1 }$. For it to be zero, we must have  $n - \mathfrak{n}_1 + 1 \geq \mathfrak{n}_2$, from which we conclude that
the nilpotency index of the tensor product
is $\n_{\rm tensor} =\mathfrak{n}_1 +  \mathfrak{n}_2 -1$.
It is immediate to show by induction that the tensor product of $k$ theories with nilpotency indices $\mathfrak{n}_1, \mathfrak{n}_2,\dots, \mathfrak{n}_k$ is
    \begin{equation}
    \label{eq:ntensor}
     \mathfrak{n}_{\text{tensor}}=\left(\sum_{i=1}^k \mathfrak{n}_i\right)-k+1\;.
    \end{equation}
We obviously have that the rank of the tensor product is 
the sum of the individual ranks,
\begin{equation}
 \mathfrak{rank}_{\rm tensor} =
  \sum_{i=1}^k \mathfrak{rank}_i \,.
 \end{equation}
 Thus if conjecture~1 holds for the individual theories,
it also holds for their tensor product,
\begin{equation}
    \mathfrak{rank}_{\rm tensor} =\sum_{i=1}^k \mathfrak{rank}_i \leq \left(\sum_{i=1}^k (\mathfrak{n}_i-1) \right)=\mathfrak{n}_{\text{tensor}}-1\,.
\end{equation}
This result will come in handy in Section~\ref{sec:RGFlow}.
\subsection{Discrete gaugings}

Here we comment briefly on the behaviour of $\mathfrak{n}$ under discrete gaugings. Given a theory $\mathcal{T}$ with  corresponding VOA $\mathcal{V}$, gauging a non-anomalous discrete group $G$ of $\mathcal{T}$ simply amounts to projecting on the fixed points of the action of $G$ on the VOA $\mathcal{V}$. We denote the VOA obtained this way as $\mathcal{V}/G$. In any VOA, there are infinitely many nulls of the form
\begin{equation}
	T^n+\phi_n,\;\; \phi_n \in C_2(\mathcal{V}),\;\;n\geq\mathfrak{n}\,.
\end{equation}
Upon discrete gauging, some of these nulls may be projected out, as we must require $\phi_n\in\mathcal{V}/G$. Therefore, the nilpotency index can only increase or remain the same under discrete gauging.

Let us illustrate this fact in two elementary examples. The $\mathbb{Z}_2$ gaugings of the free hypermultiplet and the free vector multiplet.

\paragraph{Free hyper:}
 $\mathfrak{n}=1$ for the free hyper. Let us perform a $\mathbb{Z}_2$ gauging which acts on the $\beta\gamma$ VOA as
\begin{equation}
	\mathbb{Z}_2: (\beta,\gamma)\mapsto (-\beta,-\gamma)\;.
\end{equation}
The generators of the VOA after the discrete gauging are 
\begin{equation}
	J^+=\beta^2,\;\;J^0=\beta\gamma,\;\;J^-=\gamma^2\;.
\end{equation}
These generate a $\mathfrak{su}(2)_{-\frac{1}{2}}$ affine Kac-Moody (AKM) algebra. The stress tensor is the Sugawara stress tensor. A simple calculation shows that 
 \begin{equation}
\mathfrak{n}_{\mathfrak{su}(2)_{-\frac{1}{2}}} = \mathfrak{ord}(\mathcal{D})_{\mathfrak{su}(2)_{-\frac{1}{2}}}  =3\,.
 \end{equation}
\paragraph{Free vector}
The VOA for the $\mathbb{Z}_2$ gauging of the free vector multiplet is the triplet VOA with $c=-2$ \cite{Kausch:1995py}. If one starts with the VOA of the free vector discussed in Section \ref{sec:sympferm}, which is generated by symplectic fermions $\lambda_{\alpha=1,2}$. The $\mathbb{Z}_2$ acts with a minus sign on the symplectic fermions
\begin{equation}	\mathbb{Z}_2\;:\;\lambda_\alpha\mapsto-\lambda_\alpha\;.
\end{equation}
The VOA obtained after going to the fixed point of this action is generated by a stress tensor $T$ and a triplet $W_{\alpha\beta}$ which can written in terms of the symplectic fermions as follows
\begin{equation}
	T=-\frac{1}{2}\epsilon^{\alpha\beta}\lambda_\alpha\lambda_\beta,\;\;W_{\alpha\beta}=\lambda_{(\alpha}\partial\lambda_{\beta)}\;.
\end{equation}
 We have computed the nilpotency index and order of the MLDE and found
 \begin{equation}
\mathfrak{n}_{\rm triplet} = \mathfrak{ord}(\mathcal{D})_{\rm triplet} =5\,.
 \end{equation}
 In our calculation, we found it useful to use the expression of the generators in terms of 
 free fermions:
 the nilpotency null
 is automatically zero once the generators are written in terms of fermions. Free field realizations
 will be used extensively in Section \ref{sec:Methods}.

%% file: sections/S3.tex
\section{Brief outline of methods used to find the nilpotency index}
\label{sec:Methods}

In this section we briefly describe the three different methods which we have used to calculate the nilpotency index:
\begin{enumerate}
    \item Finding the nilpotency null
    from the presentation of the VOA in terms of strong generators and their OPEs.
    \item 
    Finding the nilpotency null by
    leveraging free field realizations.
    \item Detecting indirectly the nilpotency null from the superconformal index.
\end{enumerate}
The first two methods determine the nilpotency null explicitly, while the last usually only gives a good indication of its existence. In the rest of this section we will use the example of $SU(2)$ SQCD to illustrate each method. All the computations involving the explicit determination of the null were done using the Mathematica packages \cite{thielemans1995algorithmic, Fujitsu:1994np}. We also found the notebooks of \cite{Agarwal:2021oyl} useful for performing certain computations which require the free field realization of \cite{Bonetti_2019}.
\subsection{Directly from the VOA }
In this method, the starting point is the presentation
of the VOA  in terms of strong generators and their OPEs. We systematically search for a  null of the form
$\mathfrak{N} = T^n + \phi$, $\phi \in C_2 ({\cal V})$,
for increasing values of $n$. Once we find it, we stop the search. In practice, to detect the null, we compute norms, as having vanishing norm is a necessary condition for a null. Once we find a candidate with zero norm we must also check that it is orthogonal to all other states.

For $SU(2)$ SQCD the VOA is the AKM $\mathfrak{so}(8)_{-2}$, which has generators $J^{a}$ in the $\mathbf{28}$ of $\mathfrak{so}(8)$, with the usual AKM OPEs
\begin{equation}
	J^a(z) J^b(w)\sim \frac{-2\kappa^{a b}}{(z-w)^2}+\frac{f^{a b}_{\;\;\;c} J^c}{z-w}\;.
\end{equation}
The lowest null vector of the requisite form is
\begin{equation}
	\label{eq:N2null}
	T^2-\frac{3}{10}\partial^2 T-\frac{\kappa_{ab}}{10}\left(3\partial J^a \partial J^b-2\partial^2 J^a J^b\right)\;,
\end{equation}
therefore in this case $\mathfrak{n}=2$. Clearly $\mathfrak{n}=\rank+1$.

\subsection{Free field realizations}
\label{sec:ffrtech}
Many VOAs associated to 4d $\mathcal{N}=2$ SCFTs admit generalized free field realizations\footnote{A related free field realization for VOAs corresponding to 4d $\mathcal{N}=3$ and $\mathcal{N}=4$ theories
was introduced in \cite{Bonetti_2019} and utilizes $\beta\gamma$ and $bc$ fields.} 
\cite{Beem_2019,Beem_2020,Beem:2021jnm, Beem:2024fom}. These constructions realize a given VOA in terms of building blocks associated to the low energy degrees of freedom of the Higgs Branch. The strong generators of the VOA are expressed in terms of the free fields (and possibly strongly coupled building blocks) that survive in the IR. Conjecturally, these free field constructions realize the \textit{simple quotient} of the VOA, i.e.~null vectors are identically equal to zero, when the operators are substituted by their free field expressions. This gives a very convenient algorithm to find nulls.

What's more,  the free field realizations can be used to assign $R$-charges to composite fields. Recall that Schur operators are graded by the Cartan $R$ of the $SU(2)_R$ symmetry,
but that $R$ is only a filtration at the level of the VOA structure -- it is non-increasing under normal ordered product. The $R$-filtration of the VOA (and indeed the $R$-grading of the vector space of Schur operator) is unambiguous if we can track its 4d origin.
But if we are handed the VOA abstractly, without reference to 4d (say in terms of its set of strong generators, defined below, and their singular OPEs), the $R$-filtration is a priori unknown.  This problem is (conjecturally) solved by the free realization, which comes equipped with a rule to assign $R$ to all composites.

Let us consider the example of $\mathfrak{so}(8)_{-2}$.
Its free field realization is a little involved.
We won't give the explicit expressions for the generators (see \cite{Beem_2019} for all the details) but just enumerate the requisite  free fields.
They consist of chiral bosons $\delta$, $\varphi$ and symplectic bosons $\xi_{abc}$ in the trifundamental of $\mathfrak{a}_1\oplus\mathfrak{a}_1\oplus\mathfrak{a}_1$, with the following OPEs
\begin{equation}
	\delta(z)\delta(w)\sim \,\log (z-w)~,\qquad
	\varphi(z)\varphi(w)\sim -\,\log (z-w)~,\qquad
	\delta(z)\varphi(w)\sim0~,
\end{equation}
\begin{equation}
	\xi^{a_1b_1c_1}(z)\xi^{a_2b_2c_2}(w)\sim\frac{\epsilon^{a_1a_2}\epsilon^{b_1b_2}\epsilon^{c_1c_2}}{z-w}~.
\end{equation}
As already mentioned,  the null vector~\eqref{eq:N2null} is identically zero once the generators are expressed in terms of the free fields,
so that one avoids the  computation of norms.

 \subsection{Superconformal index}
 \label{sec:superconformalindex}
Finally the superconformal index can be used to detect nulls. The superconformal index of a 4d $\mathcal{N}=2$ theory encodes protected data of the spectrum up to recombinations. By construction, it is invariant under marginal deformations. The $\mathcal{N}=2$ superconformal index is given by
 \begin{equation*}
 	\begin{split}
 		\mathcal{I}=\text{Tr}(-1)^Fp^{\frac{1}{2}(E-2j_1-2R-r)}q^{\frac{1}{2}(E+2j_1-2R-r)}t^{R+r}\;,
 	\end{split}
 \end{equation*}
 where $p,q$ and $t$ are superconformal fugacities and this index can be further refined by flavour fugacities. There are different limits of the index but we will focus mainly on Schur and Macdonald limit.
 
 \textbf{Schur limit:} $t=q$ with $p$ arbitrary,
 \begin{equation}
 	\mathcal{I}_{\text{Schur}}(q)=\text{Tr}(-1)^F q^{E-R}= \text{Tr}(-1)^F q^{h}\;.
 \end{equation}
 
 \textbf{Macdonald limit:} $p\to0$ and $q,t$ arbitrary,
 \begin{equation}
 	\label{eq:macindex}
 	\mathcal{I}_{\text{Mac}}(q,t)=\text{Tr}(-1)^F q^{E-2R-r} t^{R+r}   \;.
 \end{equation}
 In fact we will find it more useful to perform the change of variables $t=q \TMac$,
  \begin{equation} 	\label{eq:macrenamvar}	\mathcal{I}_{\text{Mac}} (q, \TMac) =\text{Tr}(-1)^F q^{E-R} \TMac^{R+r}=\text{Tr}(-1)^F q^{h} \TMac^{R+r}~. 
 \end{equation}
As the Macdonald index contains information
about the crucial $R$ quantum number, it 
 is better suited for detecting nulls of the requisite form.  
The stress tensor is canonically assigned $R=1$. The nilpotency null is of the form
 \begin{equation}
 	\label{eq:nullvector}	   \mathcal{N}=T^n+\phi,\;\;\;\;\;\phi\in C_2(\mathcal{V})\;.
 \end{equation}
 This state has $h=2n$. The
 term $T^n$ should have  $R$-assignment $n$, while $\phi$ may contain terms with $R \leq n$. 

 A convenient tool to determine generators and relations from the superconformal index is the \textit{plethystic logarithm} (PLog),
 \begin{equation}  \text{PLog}[f(x,y,...)]=\sum_{l=1}^{\infty}\frac{\mu(l)}{l}\ln f(x^l,y^l,...)\;,
 \end{equation}
 where $\mu$ is the M\"obius function. 
 From the $q$-expansion of the plethystic log of the Macdonald index we can (tentatively) read off the strong generators and the nulls, at least for low powers of $q$. Terms with positive integer coefficients at low orders are identified with bosonic generators while terms with negative integer coefficients can be identified either with fermionic generator or relations/null states. It is  convenient to organize the terms in $SL(2)$ families and consider only the contributions of quasi-primaries. This is simply accomplished by multiplying the plethystic logarithm by $(1-q)$ which takes into account derivative operators. In summary, the existence of a nilpotency null of the form \eqref{eq:nullvector} is indicated by the presence of the following term in the plethystic logarithm of the Macdonald index
 \begin{equation}
 \label{MacCrit}
    (1-q)\,\text{PLog}(\mathcal{I}_{\text{Mac}}) 
    \supset -q^{2n}\TMac^n\;.
 \end{equation}
 Some caveats are in order. 
 In principle the term (\ref{MacCrit}) may come from a state of the form \eqref{eq:nullvector} but with $\phi\notin  C_2(\mathcal{V})$.  There may also be several cancellations in the index that prevent one from unambiguously detecting the nilpotency null. Flavor symmetries help, as one can often refine the index by flavor fugacities, 
 and restrict to singlets. All in all, the Macdonald index
 can give a very good indication for the existence
 of the nilpotency null, while often falling short of a complete proof.

 Let us see how this works in the case of $SU(2)$ SQCD. The Macdonald index reads
 \begin{equation}
 	\begin{split}
 		\mathcal{I}^{SU(2)\,\text{SQCD}}_{Mac}&=1+28 q \TMac+q^2 \TMac (29+300 \TMac)+q^3 \TMac \left(29+678 \TMac+1925 \TMac^2\right)\\&+q^4 \TMac \left(29+1112 \TMac+6321 \TMac^2+8918 \TMac^3\right)+\dots\;
 	\end{split}
 \end{equation}
and its plethystic log is
 \begin{equation}
 	\begin{split}
 		\text{PLog}(\mathcal{I}^{SU(2)\,\text{SQCD}}_{Mac})&=\frac{28 q \TMac + q^2 (\TMac - 106 \TMac^2) + q^3 (-28 \TMac^2 + 833 \TMac^3)}{1-q}\\& 
 		+\frac{ 
 			q^4 (-\redTMac^2 + 540 \TMac^3 - 8400 \TMac^4)+\dots}{1-q}\;.
 	\end{split}
 \end{equation}
 The $-q^4\redTMac^2$ term indicates the presence of a nilpotency null with $\mathfrak{n}=2$. This is the state \eqref{eq:N2null}. The  MLDE for the vacuum character of this theory is of second order.
In summary,
 \begin{equation}
 	\text{ord}(\mathcal{D})=\mathfrak{n}=\rank+1=2
 \end{equation}
consistent  with our conjecture~1.

%% file: sections/Sresults.tex
\section{Results}
\label{sec:resultsdiscuss}
In this section we briefly describe our concrete calculations.
The results are summarized in table \ref{tab:summaryres}.
Whenever the plethystic logarithm of the Macdonald index is shown, the term which is supposed to indicate the presence of the nilpotency null is highlighted in \textcolor{red}{red},
$-q^{2\ioni} \redTMac^\ioni$. 
\subsection{$\mathcal{N}=2$ SQCD and $\mathcal{N}=4$ SYM}
The most elementary theories that one would want to study are Lagrangian theories. We consider $\mathcal{N}=2$ SQCD and $\mathcal{N}=4$ SYM, with gauge groups $SU(2)$ and $SU(3)$.

The only rank-one theories which are Lagrangian are $SU(2)$ $\mathcal{N}=2$ SQCD and $SU(2)$ $\mathcal{N}=4$ SYM. The associated VOA for $SU(2)$ SQCD is the AKM $\mathfrak{so}(8)_{-2}$ and the VOA for $\mathcal{N}=4$ $SU(2)$ SYM is the small $\mathcal{N}=4$ superconformal algebra with $k=-\frac{3}{2}$. $\ioni$ for both of these theories is $2$ and the Schur index is annihilated by an order $2$ MLDE. 

For $SU(3)$ SQCD, finding the explicit null is difficult and here we try to detect it from the Macdonald index, 
\begin{equation}
	\begin{split}
		\text{PLog}(\mathcal{I}^{SU(3) \text{SQCD}}_{\text{Mac}})=&\frac{36 \TMac q+40 \TMac^{3/2} q^{3/2}+\left(\TMac-36 \TMac^2\right) q^2-320 \TMac^{5/2} q^{5/2}-435 \TMac^3 q^3}{1-q}\\
		+&\frac{\left(-40 \TMac^{5/2}+1944 \TMac^{7/2}\right)
		q^{7/2}+\left(-226 \TMac^3+8627 \TMac^4\right) q^4+\left(360 \TMac^{7/2}-264 \TMac^{9/2}\right) q^{9/2}}{1-q}\\
		+&\frac{\left(-36 \TMac^3+5461 \TMac^4-96040
		\TMac^5\right) q^5+\left(8176 \TMac^{9/2}-223536 \TMac^{11/2}\right) q^{11/2}}{1-q}\\
		+&\frac{\left(\textcolor{red}{-\TMac}^3+1623 \TMac^4-71978 \TMac^5+585612 \TMac^6\right) q^6+\dots}{1-q}\;.
	\end{split}
\end{equation}
The term $-q^6\redTMac^3$ indicates that there might exist a nilpotency null and $\ioni=3$ for this theory.

The Macdonald index for $SU(3)$ $\mathcal{N}=4$  SYM has a lot of cancellations and the null is not visible in the Macdonald index. \cite{Agarwal:2021oyl} provide a Mathematica notebook which computes the superconformal indices using free field for the VOA for $\mathcal{N}=3$ SCFTs, which we adapt for $SU(3)$ $\mathcal{N}=4$ SYM. The free fields have definite quantum number assignments and therefore the character of VOA can be refined more. They include two fugacities $\xi$ and $\nu$ such that the character computed by their notebooks evaluates $\text{Tr}(-1)^F q^{E-R} \xi^{R}\nu^{r}$, which we will denote by $\chi$ to avoid confusion with the index (see a more detailed discussion in \cite{Bonetti_2019}). The presence of the nilpotency null is detected by a term like $-\frac{q^{2n} \xi^n}{1-q}$ in the plethystic logarithm. Using the free fields for $SU(3)$ $\mathcal{N}=4$  SYM, we get the following expression where the $-\frac{q^6\xi^3}{1-q}$ term indicates presence of a nilpotency null such that $\ioni=3$.
\begin{equation}
	\begin{split}
			\text{PLog}(\chi^{\mathcal{N}=4\,  SU(3)}_{\text{Mac}})=&\frac{3 q \xi+q^{3/2} \left(4 \xi^{3/2}-4 \xi\right)+q^2 \left(\xi-6 \xi^{3/2}\right)+q^{5/2} \left(2 \xi^{3/2}-2 \xi^{5/2}\right)}{1-q}\\+ &\frac{q^3 \left(8 \xi^{5/2}-3 \xi^3\right)+q^{7/2} \left(12 \xi^3-12 \xi^{5/2}\right)+q^4 \left(3 \xi^4-2 \xi^{7/2}-19 \xi^3+8 \xi^{5/2}\right)}{1-q}\\+&\frac{q^{9/2} \left(2 \xi^{9/2}-24 \xi^4+8 \xi^{7/2}+16 \xi^3-2 \xi^{5/2}\right)}{1-q}\\+&\frac{q^5 \left(-\xi^5-24 \xi^{9/2}+70 \xi^4-12 \xi^{7/2}-9 \xi^3\right)}{1-q}\\+&\frac{q^{11/2} \left(-6 \xi^{11/2}+16 \xi^5+82 \xi^{9/2}-104 \xi^4+8 \xi^{7/2}+4 \xi^3\right)}{1-q}\\+&\frac{q^6 \left(-3 \xi^6+66 \xi^{11/2}-72 \xi^5-136 \xi^{9/2}+94 \xi^4-2 \xi^{7/2}-\textcolor{red}{\xi}^3\right)+\dots}{1-q}\;.
		\end{split}
\end{equation}

\subsection{Deligne rank-one theories}
\label{sec:drankone}
VOAs corresponding to Deligne rank-one theories are AKM VOAs based on the \textit{Deligne-Cvitanovic exceptional series} of simple Lie algebras \footnote{$\mathfrak{g}_2$ and $\mathfrak{f}_4$ entries of the series are not known to correspond to existing rank-one SCFTs.}
\begin{equation}
	\mathfrak{a}_0\subset\mathfrak{a}_1\subset\mathfrak{a}_2\subset\mathfrak{g}_2\subset\mathfrak{d}_4\subset\mathfrak{f}_4\subset\mathfrak{e}_6\subset\mathfrak{e}_7\subset\mathfrak{e}_8\;,
\end{equation}
with level $k=-\frac{h^\vee}{6}-1$, where $h^\vee$ denotes the dual Coxeter number. The OPEs are that of AKM algebra,
\begin{equation}
	J^a(z) J^b(w)\sim \frac{k\kappa^{a b}}{(z-w)^2}+\frac{f^{a b}_{\;\;\;c} J^c}{z-w}\;.
\end{equation}
The stress tensor for these theories is the Sugawara stress tensor $T=\frac{\kappa_{ab}}{2(k+h^\vee)}J^aJ^b$. The nilpotency index for these theories is $2$ and the nilpotency null has the following explicit form
\begin{equation}
	T^2-\frac{3}{10}\partial^2 T+\frac{3\kappa_{ab}}{5(h^\vee+6k)}\left(3\partial J^a \partial J^b-2\partial^2 J^a J^b\right)
\end{equation}
\subsection{Argyres-Douglas theories}
In this section we discuss Argyres-Douglas theories \cite{Argyres:1995jj, Xie:2012hs} of type $(A_1,G)$ for $G=A_{n}, D_{n}, E_6, E_8$, $n\in\mathbb{Z}_{>0}$. $G=A_n, D_n$ theories were discussed in \cite{Beem_2018} from the point of view of MLDEs. Nilpotency index was also mentioned for most of the theories but not all in that reference. Due to computational limitations we were unable to determine $\mathfrak{n}$ for the case $G=E_7$.
The Macdonald indices for $G=A_n,D_n$ were conjectured in  \cite{Song:2015wta, Buican:2015tda}.

\subsubsection{$(A_1,A_{2n})$}
\label{sec:A1A2n}
The VOA of $(A_1,A_{2n})$ theory is Virasoro VOA $(2,2n+3)$ with central charge $c=-\frac{2n(6n+5)}{2n+3}$.  These theories have $\mathfrak{rank}=n$. Any $(p,q)$ minimal model has a null vector of the form $T^{\frac{1}{2}(p-1)(q-1)}$ plus operators with at least one derivative. Therefore for $(A_1,A_{2n})$,  $\ioni=\frac{(1)(2n+2)}{2}=n+1$. As these VOAs are just Virasoro, writing the nulls is very straightforward. The nulls for $n=1,2$ and $3$ are as follows
\begin{align}
 (A_1,A_2):	&\, T^2-\frac{3}{10}\partial^2T\;, \\
 (A_1,A_4): &\, T^3-\frac{1}{7} \partial T\partial T-\frac{11}{14} \partial^2 T\,T-\frac{19}{588}\partial^4T\;,\\
 (A_1,A_6): &\,	T^4-\frac{4}{9}\partial T\partial T T-\frac{13}{9}\partial^2 T T^2+\frac{7}{36} \partial^2 T\partial^2T+\frac{2}{81}				\partial^3T\partial T-\frac{11}{81} \partial^4 T T-\frac{139}{29160}\partial^6T\;.
\end{align}

We can also detect the nulls from the Macdonald index of these theories. We show the plethystic logarithm for $n=1,2$ and $3$. The terms containing $T$ in red, indicate the presence of nilpotency null,
\begin{align}
	&\text{PLog}(\mathcal{I}^{(A_1,A_2)}_{\text{Mac}})=\frac{q^2 \TMac-q^4 \redTMac^2-q^8 \TMac^2+\dots}{1-q}\;,\\
	&\text{PLog}(\mathcal{I}^{(A_1,A_4)}_{\text{Mac}})=\frac{q^2 \TMac-q^6 \redTMac^3+q^7 \left(-\TMac-2 \TMac^2\right)+\dots}{1-q}\;,\\
	&\text{PLog}(\mathcal{I}^{(A_1,A_6)}_{\text{Mac}})=\frac{q^2 \TMac-q^8 \redTMac^4+q^9 \left(-\TMac-3 \TMac^2-3 \TMac^3\right)+\dots}{1-q}\;.
\end{align}
As expected, for $(A_1,A_{2n})$ theories, we can see that the term of the form $-q^{2(n+1)}\redTMac^{n+1}$ occurs in the plethystic logarithm of the Macdonald index.

\subsubsection{$(A_1,D_{2n+1})$}
The VOA of $(A_1,D_{2n+1})$ is the affine Kac-Moody algebra $\mathfrak{su}(2)_{-\frac{4n}{2n+1}}$. These theories have rank $n$. For $n=1,2$ we have found nulls, such that $\ioni$ is $2$ and $3$ respectively. Let us also note the plethystic logarithms for $n=1,2$ and $3$.
\begin{align}
	&\text{PLog}(\mathcal{I}^{(A_1,D_3)}_{\text{Mac}})=\frac{3 q \TMac-3 q^3 \TMac^2+q^2 \left(\TMac-\TMac^2\right)+q^4 \left(-\redTMac^2+4 \TMac^3\right)+q^5 \left(4 \TMac^3-4 \TMac^4\right)+\dots}{1-q}\;,\\
	&\text{PLog}(\mathcal{I}^{(A_1,D_5)}_{\text{Mac}})=\frac{3 q \TMac-3 q^5 \TMac^3+q^2 \left(\TMac-\TMac^2\right)+q^6 \left(-\redTMac^3+4 \TMac^4\right)+\dots}{1-q}\;,\\
	&\text{PLog}(\mathcal{I}^{(A_1,D_7)}_{\text{Mac}})=\frac{3 q \TMac-3 q^7 \TMac^4+q^2 \left(\TMac-\TMac^2\right)+q^8 \left(-\redTMac^4+4 \TMac^5\right)+\dots}{1-q}\;.
\end{align}

\subsubsection{$(A_1,A_{2n+1})$}
The VOA is the subregular DS-reduction of the $\mathfrak{su}(n+1)_{-\frac{(n+1)^2}{n+2}}$ theory and is called $\mathcal{B}_{n+2}$ algebra. By explicit determination of nulls we have found $\ioni$ for $n=1,2,3$ and it equals $2,3$ and $4$ respectively. The Macdonald indices for $n=1,2,3$ are as follows

\begin{equation}
	\text{PLog}(\mathcal{I}^{(A_1,A_3)}_{\text{Mac}})=\frac{3 \TMac q+\left(\TMac-\TMac^2\right) q^2-3 \TMac^2 q^3+\left(-\redTMac^2+4 \TMac^3\right) q^4+\dots}{1-q}\;,
\end{equation}
\begin{equation}
	\begin{split}
	\text{PLog}(\mathcal{I}^{(A_1,A_5)}_{\text{Mac}})&=\frac{\TMac q+2 \TMac^{3/2} q^{3/2}+\TMac q^2-\TMac^3 q^3-2 \TMac^{5/2} q^{7/2}-\TMac^3 q^4}{1-q}\\
		&+\frac{\left(-\TMac^3+2 \TMac^4\right) q^5+2 \TMac^{9/2}
			q^{11/2}+\left(-\redTMac^3+2 \TMac^4\right) q^6+\dots}{1-q}\;,
	\end{split}
\end{equation}
\begin{equation}
	\begin{split}
	\text{PLog}(\mathcal{I}^{(A_1,A_7)}_{\text{Mac}})&=\frac{\TMac q+\left(\TMac+2 \TMac^2\right) q^2+\left(-2 \TMac^3-\TMac^4\right) q^4-\TMac^4 q^5+\left(-\TMac^4+2 \TMac^5\right) q^6}{1-q}\\
	&+\frac{\left(-\TMac^4+2 \TMac^5+2
		\TMac^6\right) q^7+\left(-\redTMac^4+2 \TMac^5+\TMac^6-2 \TMac^7\right) q^8+\dots}{1-q}\;.
	\end{split}
\end{equation}

\subsubsection{$(A_1,D_{2n+2})$}
The VOA is the sub-subregular DS-reduction of the
$\mathfrak{su}(n+2)_{-\frac{n^2+2n}{n+1}}$ theory and is called the $\mathcal{W}_{n+1}$ algebra. By explicit determination of nulls we have found $\ioni$ for $n=1,2$ and it equals $2$ and $3$  respectively. The Macdonald indices for $n=1,2,3$ are as follows

\begin{equation}
	\text{PLog}(\mathcal{I}^{(A_1,D_4)}_{\text{Mac}})=\frac{8 \TMac q+\left(\TMac-9 \TMac^2\right) q^2+\left(-8 \TMac^2+16 \TMac^3\right) q^3+\left(-\redTMac^2+38 \TMac^3-45 \TMac^4\right) q^4+\dots}{1-q}\;,
\end{equation}
\begin{equation}
	\begin{split}
	\text{PLog}(\mathcal{I}^{(A_1,D_6)}_{\text{Mac}})=&\frac{4 \TMac q+4 \TMac^{3/2} q^{3/2}+\left(\TMac-\TMac^2\right) q^2-4 \TMac^{5/2} q^{5/2}-4 \TMac^3 q^3+4 \left(-\TMac^{5/2}+\TMac^{7/2}\right) q^{7/2}}{1-q}\\
	&+\frac{2\left(-2 \TMac^3+4 \TMac^4\right) q^4+4 \TMac^{7/2} q^{9/2}-2 \left(2 \TMac^3-9 \TMac^4+7 \TMac^5\right) q^5}{1-q}\\
	&+\frac{-4 \left(-3 \TMac^{9/2}+4
		\TMac^{11/2}\right) q^{11/2}+\left(-\redTMac^3+16 \TMac^4-36 \TMac^5+17 \TMac^6\right) q^6+\dots}{1-q}\;,
	\end{split}
\end{equation}
\begin{equation}
	\begin{split}
	\text{PLog}(\mathcal{I}^{(A_1,D_8)}_{\text{Mac}})=&\frac{4 \TMac q+\left(\TMac+3 \TMac^2\right) q^2-4 \TMac^3 q^3-4 \TMac^3 q^4+4 \TMac^5 q^5+\left(-4 \TMac^4+14 \TMac^5-6 \TMac^6\right) q^6}{1-q}\\
		&+\frac{\left(-4 \TMac^4+16
			\TMac^5-16 \TMac^6\right) q^7+\left(-\redTMac^4+16 \TMac^5-14 \TMac^6-20 \TMac^7+15 \TMac^8\right) q^8+\dots}{1-q}\;.
	\end{split}
\end{equation}

\subsubsection{$(A_1,E_6)$, $(A_1,E_8)$}

The VOAs for $(A_1,E_6)$ and $(A_1,E_8)$ are $W_3(3,7)$ and $W_3(3,8)$ respectively. The theories $(A_1,E_6)$ and $(A_1,E_8)$ are isomorphic to $(A_2,A_3)$ and $(A_2, A_4)$ respectively. These are $W_3$ algebras with central charges $-\frac{114}{7}$ and $-23$. The OPEs of $W_3$ algebra are as follows
\begin{equation}
	\begin{split}
		T(z)T(w)&\sim \frac{\frac{c}{2}}{(z-w)^4}+\frac{2 T(w)}{(z-w)^2}+\frac{\partial T(w)}{z-w}\;,\\
		T(z)W(w)&\sim \frac{3W(w)}{(z-w)^2}+\frac{\partial W(w)}{z-w}\;,\\
		W(z)W(w)&\sim \frac{\frac{c}{3}}{(z-w)^6}+\frac{2T(w)}{(z-w)^4}+\frac{\partial T(w)}{(z-w)^3}+\frac{\frac{32}{22+5c}\left(T(w)^2-\frac{3}{10}\partial^2T\right)+\frac{3}{10}\partial^2T}{(z-w)^2}\\
		&+\frac{\frac{16}{22+5c}\partial\left(T(w)^2-\frac{3}{10}\partial^2T\right)+\frac{1}{15}\partial^3T}{z-w}\;.
	\end{split}
\end{equation}
The nilpotency nulls are as follows
\begin{equation}
	(A_1,E_6) : 
	T^4-\frac{16}{35}(\partial T)^2T-\frac{26}{35}(\partial W)^2-\frac{53}{35}(\partial^2 T)\,T^2+\frac{27}{140}\partial^2 T\partial^2 T+\frac{208}{245}\partial^2 W\,W-\frac{41}{294}(\partial^4 T)\,T-\frac{169}{44100}\partial^6 T\;,
\end{equation}
\begin{equation}
	\begin{split}
		(A_1,E_8) : 
		&T^5-\frac{9765}{1664}T\partial W\partial W+\frac{4185}{832}T\partial^2 W\,W-\frac{45}{52}\partial T\partial T T^2\\
		&-\frac{30}{13}\partial^2 T T^3+\frac{27}{104}\partial^2 T\partial T\partial T+\frac{2367}{3328}\partial^2 T\partial^2 T T-\frac{2511}{3328}\partial^2 W\partial^2 W\\
		&-\frac{147}{1664}\partial^3 T\partial T T-\frac{261}{6656}\partial^3 T\partial^3 T+\frac{11439}{6656}\partial^3 W\partial W-\frac{659}{1664}(\partial^4 T)\,T^2\\
		&+\frac{219}{6656}\partial^4 T\partial^2 T-\frac{9393}{13312}\partial^4 W W-\frac{3189}{133120}\partial^5 T\partial T-\frac{1501}{66560}(\partial^6 T)\,T+\frac{129}{266240}\partial^8 T\;.
	\end{split}
\end{equation}
Therefore $\ioni=4$ and $5$ respectively. In these examples we would like to point that there are other nulls which are very similar to the nilpotency null and may contribute $\frac{-q^{2n}\TMac^n}{1-q}$ to the index but one would reach the wrong conclusion in doing so. Example of the nulls of similar form but not nilpotency nulls in these theories are as follows

\begin{equation}
	(A_1,E_6):
	T^3-\frac{39}{7}W^2-\frac{6}{7}\partial T\partial T-\frac{12}{7}\partial^2T\, T+\frac{17}{196}\partial^4 T\;,
\end{equation}
\begin{equation}
	(A_1,E_8):
	T^3-\frac{279}{16}W^2-\frac{51}{32}\partial T\partial T-\frac{21}{8}\partial^2T\, T+\frac{9}{32}\partial^4 T\;.
\end{equation}
\subsection{Rank-one $\mathcal{N}=3$ theories}
\label{sec:n3theories}
Another interesting class of theories is the rank-one $\mathcal{N}=3$ theories \cite{Aharony:2015oyb, Garcia-Etxebarria:2017ffg, Nishinaka:2016hbw}. These theories are labelled by an integer $\ell$, with $\ell=2,3,4,6$  and the moduli space of these theories is $\mathbb{C}^3/\mathbb{Z}_\ell$.  Because of additional supersymmetry, the definition of Higgs branch and Coulomb branch depends on the choice of the $\mathcal{N}=2$ inside $\mathcal{N}\geq 3$ superconformal subalgebra used to define the branches. The VOA of these theories has central charge $c_{2d}=3-6\ell$ and is generated by the operators $\langle\mathcal{T},\mathcal{J},\mathcal{G},\tilde{\mathcal{G}},\mathcal{W},\tilde{\mathcal{W}},\mathcal{L},\tilde{\mathcal{L}}\rangle$
with dimension $2,1,\frac{3}{2},\frac{3}{2},\frac{\ell}{2},\frac{\ell}{2},\frac{\ell+1}{2}$ and $\frac{\ell+1}{2}$ respectively. $\mathcal{T}$ is the stress tensor and is not an independent generator when $\ell=2$. $\mathcal{T},\mathcal{J},\mathcal{G}$ and $\tilde{\mathcal{G}}$ generate a subalgebra, which is called the $\mathcal{N}=2$ superconformal algebra. The case of $\ell=2$ is the same as the $\mathcal{N}=4$ SYM. For $\ell=3$ and $\ell=4$, we determined $\ioni=3$ and $\ioni=4$ respectively. The $\ell=6$ case was not determined due to large computational time. 
\subsection{$A_1$ class $S$ theories with genus zero}
Another interesting set of examples are the class $S$ theories of type $A_1$ with genus zero and $s$ punctures \cite{Gaiotto:2009we}. These theories have rank $s-3$. For these theories, the method of explicitly computing the nulls is very difficult since the VOAs for these theories become quite complicated with increasing number of punctures. To get some idea for the value for $\ioni$, we use Macdonald index. Here, we show the plethystic logarithm for $s=3,4,5$ and $6$.

\begin{equation}
	\text{PLog}(\mathcal{I}^{s=3}_{\text{Mac}})=\frac{8 q^{1/2}\TMac^{1/2}}{1-q}\;,
\end{equation}
\begin{equation}
	\text{PLog}(\mathcal{I}^{s=4}_{\text{Mac}})=\frac{28 q \TMac + q^2 (\TMac - 106 \TMac^2) + q^3 (-28 \TMac^2 + 833 \TMac^3) +
		q^4 (-\redTMac^2 + 540 \TMac^3 - 8400 \TMac^4)+\dots}{1-q}\;,
\end{equation}
\begin{equation}
	\begin{split}
	\text{PLog}(\mathcal{I}^{s=5}_{\text{Mac}})&=\frac{15 q \TMac+32 q^{3/2} \TMac^{3/2}-128 q^{5/2} \TMac^{5/2}-285 q^3 \TMac^3+q^2 \left(\TMac-4 \TMac^2\right)}{1-q}\\&+
	\frac{q^{7/2} \left(-32 \TMac^{5/2}+320
		\TMac^{7/2}\right)+q^4 \left(-91 \TMac^3+2719 \TMac^4\right)+q^{9/2} \left(128 \TMac^{7/2}+3520 \TMac^{9/2}\right)}{1-q}\\
		&+\frac{q^5 \left(-15
		\TMac^3+1745 \TMac^4-14048 \TMac^5\right)+q^{11/2} \left(3360 \TMac^{9/2}-61440 \TMac^{11/2}\right)}{1-q}\\
		&+\frac{q^6 \left(-\redTMac^3+530 \TMac^4-12756
			\TMac^5-20985 \TMac^6\right)+\dots}{1-q}\;,
	\end{split}
\end{equation}

\begin{equation}
	\begin{split}
	\text{PLog}(\mathcal{I}^{s=6}_{\text{Mac}})=&\frac{18 q \TMac-320 q^3 \TMac^3+q^2 \left(\TMac+59 \TMac^2\right)+q^4 \left(-64 \TMac^3-391 \TMac^4\right)+q^5 \left(-238 \TMac^4+13137 \TMac^5\right)}{1-q}\\+&\frac{q^6
		\left(-136 \TMac^4+9522 \TMac^5-43760 \TMac^6\right)+q^7 \left(-18 \TMac^4+3759 \TMac^5-35971 \TMac^6-395120 \TMac^7\right)}{1-q}\\+&\frac{q^8 \left(-\redTMac^4+924
		\TMac^5-9879 \TMac^6-594699 \TMac^7+3998900 \TMac^8\right)+\dots}{1-q}\;.
	\end{split}
\end{equation}
Thus, indicating that $\mathfrak{n}$ can be $2,3,$ and $4$ for $s=4,5$ and $6$ respectively, consistent with our conjecture. $s=3$
corresponds to the trifundamental free hyper, whose VOA is just symplectic bosons $\xi^{abc}$, where $abc$ are the trifundamental $SU(2)$ index. The OPEs can be presented as
\begin{equation}
	\xi^{a_1b_1c_1}(z)\xi^{a_2b_2c_2}(0)\sim\frac{\epsilon^{a_1a_2}\epsilon^{b_1b_2}\epsilon^{c_1c_2}}{z}\;.
\end{equation}
The rank of this theory is zero and $\mathfrak{n}$ is $1$ because the stress tensor is in $C_2(\mathcal{V})$. Since the stress tensor is a composite, the term $-\frac{q^2\TMac}{1-q}$ is not visible in the plethystic logarithm. $s=4$ is the class S description of $SU(2)$ $\mathcal{N}=2$ SQCD and the nilpotency index is $2$.

\subsection{Deligne rank-two $\mathfrak{d}_4$}
Another theory where we try to detect the null using the Macdonald index is the Deligne rank-two $\mathfrak{d}_4$ theory \cite{Beem_2020}. This theory admits a Lagrangian description as an $\mathfrak{usp}(4)$ gauge theory with four hypers in the fundamental ($\mathbf{4}$) and one hyper in the antisymmetric ($\mathbf{5}$) of $\mathfrak{usp}(4)$. The plethystic log of the Macdonald index is given below and the $-\frac{q^6\redTMac^3}{1-q}$ indicates that $\ioni$ might be $3$.
\begin{equation}
	\begin{split} 	
		\text{PLog}(\mathcal{I}^{\tiny \text{Deligne rank-two} \mathfrak{d}_4}_{\text{Mac}})=&\frac{31 q \TMac+56 q^{3/2} \TMac^{3/2}-752 q^3 \TMac^3+q^2 \left(\TMac-\TMac^2\right)+q^{5/2} \left(2 \TMac^{3/2}-268 \TMac^{5/2}\right)}{1-q}\\+&\frac{q^{7/2}
		\left(-58 \TMac^{5/2}+268 \TMac^{7/2}\right)+q^4 \left(-218 \TMac^3+7732 \TMac^4\right)}{1-q}\\+&\frac{q^{9/2} \left(-2 \TMac^{5/2}+58 \TMac^{7/2}+19600
		\TMac^{9/2}\right)+q^5 \left(-31 \TMac^3+4049 \TMac^4-25165 \TMac^5\right)}{1-q}\\+&\frac{q^{11/2} \left(2 \TMac^{7/2}+14344 \TMac^{9/2}-288978
		\TMac^{11/2}\right)}{1-q}\\
      +&\frac{q^6 \left(-\redTMac^3+1114 \TMac^4-13867 \TMac^5-563671 \TMac^6\right)}{1-q}
	\end{split}
\end{equation}

\subsection{Rank-one \texorpdfstring{$IV^*_{Q=1}$}{} theory}
\label{sec:IVrankone}

In this subsection we analyze the
 interesting case of the putative rank-one $IV^*_{Q=1}$ theory, which shows up in the classification of rank-one Coulomb branch geometries~\cite{Argyres:2015ffa, Argyres:2015gha,Argyres:2016xmc,Argyres:2016xua,Argyres:2020nrr}. We will first make some general observations and then 
 connect to the main narrative of this paper.

 The status of this theory is not completely settled, as a consistent Coulomb branch geometry is a necessary but not sufficient condition for the full SCFT to exist. It has been
conjectured \cite{Argyres:2015ffa,Argyres:2016yzz,Argyres:2016xmc} that this theory corresponds to the IR fixed point of an RG flow stemming from the rank-one $\mathcal{N}=3$ theory with $\mathbb{C}^3/\mathbb{Z}_4$ moduli space. 
The putative theory has a  rank-one Coulomb branch, generated by a chiral operator of  $\Delta=3$, while its Higgs branch is trivial.  Its central charges  are $c_{\text{4d}}=\frac{25}{24}$ and $a_{\text{4d}}=\frac{55}{48}$.  It is then very interesting to ask whether we can find a VOA consistent with these data. 

The absolutely minimal guess, namely the simple quotient of the Virasoro VOA at $c_{2d} = -25/2$, does not work,
because it is not $C_2$-cofinite and thus incompatible with the Higgs branch conjecture. Interestingly, at this value of the central charge there is a known extension of the Virasoro VOA
to a $W(2, 7)$ algebra~\cite{blumenhagen1991w},  which is strongly generated  by the stress tensor and by an additional generator of dimension 7. This VOA is however also {\it not}
$C_2$-cofinite. Two additional extensions look much more promising: both the triplet algebra $\mathcal{W}(4)$ \cite{Kausch:2000fu, Kausch:1990vg, Creutzig:2013hma, Creutzig:2016fms} and  the doublet algebra $\mathcal{A}(4)$ \cite{feigin2011characters,feigin2007fermionic} {\it are} $C_2$-cofinite and thus legitimate candidates.

They are special instances of the two well-studied families of VOA,
labelled by a positive integer $p$:
the triplet algebras $\mathcal{W}(p)$, with strong generators of dimensions $(2, 2p-1, 2p-1, 2p-1 )$ 
and  the doublet algebras $\mathcal{A}(p)$, with strong generators of dimensions $(2, \frac{3p-2}{4}, \frac{3p-2}{4})$.
Their central charges coincide
\begin{equation}
	c (p)=13-6p-\frac{6}{p}\; ,
\end{equation}
in fact $\mathcal{A}(p)$ is an extension
of $\mathcal{W}(p)$. 
The irreducible $\mathcal{W}(p)$-modules are organized into two families,
$\Lambda(s)$ and $\Pi(s)$, with $1\leq s\leq p$. Their characters are known in closed form \cite{Fuchs:2003yu}:~\footnote{Here,
	\begin{equation}
		\eta(q)
		=
		q^{1/24}\prod_{n=1}^{\infty}(1-q^n),
	\end{equation}
	and the level-$p$ theta functions are defined by
	\begin{equation}
		\theta_{s,p}(q,z)
		=
		\sum_{j\in\mathbb{Z}+\frac{s}{2p}}
		q^{p j^2}z^j,
		\qquad
		\theta_{s,p}(q)
		=
		\theta_{s,p}(q,1),
	\end{equation}
	with
	\begin{equation}
		\theta'_{s,p}(q)
		=
		\left.
		z\frac{\partial}{\partial z}
		\theta_{s,p}(q,z)
		\right|_{z=1}.
	\end{equation}
}
\begin{align}
	\chi^{\Lambda}_{s,p}(q)
	&=
	\frac{1}{\eta(q)}
	\left(
	\frac{s}{p}\,\theta_{p-s,p}(q)
	+
	2\,\theta'_{p-s,p}(q)
	\right),
	\nonumber\\[2mm]
	\chi^{\Pi}_{s,p}(q)
	&=
	\frac{1}{\eta(q)}
	\left(
	\frac{s}{p}\,\theta_{s,p}(q)
	-
	2\,\theta'_{s,p}(q)
	\right),
	\qquad
	1\leq s\leq p.
	\label{eq:Wp-characters}
\end{align}
The vacuum module of $\mathcal{W}(p)$ is $\Lambda(1)$. Moreover, regarded as a $\mathcal{W}(p)$-module, the vacuum module of the extended algebra $\mathcal{A}(p)$ decomposes as
\begin{equation}
	\mathcal{A}(p)
	\simeq
	\Lambda(1)\oplus\Pi(1).
\end{equation}
For $p=4$, the vacuum characters take the form
\begin{align}
	\chi_{\mathcal{W}(4)}(q)
	&=
	\chi^{\Lambda}_{1,4}(q)
	=
	\frac{1}{\eta(q)}
	\left(
	\frac{1}{4}\theta_{3,4}(q)
	+
	2\theta'_{3,4}(q)
	\right),\\
	\chi_{\mathcal{A}(4)}(q)
	&=
	\chi^{\Lambda}_{1,4}(q)
	+
	\chi^{\Pi}_{1,4}(q)
	=
	\frac{1}{\eta(q)}
	\left(
	\frac{1}{4}
	\left(
	\theta_{3,4}(q)+\theta_{1,4}(q)
	\right)
	+
	2
	\left(
	\theta'_{3,4}(q)-\theta'_{1,4}(q)
	\right)
	\right).
	\label{eq}
\end{align}
Modular transformations of the characters are known explicitly \cite{Fuchs:2003yu}. Applying these transformations to the vacuum characters of $\mathcal{A}(4)$  and
$\mathcal{W}(4)$ shows that 
in both cases, their modular $S$-transforms contain a leading 
contribution associated with chiral weight $h=-5/16$.
This contribution reproduces the expected $e^{-\frac{4\pi i}{\tau}(c_{\mathrm{4d}}-a_{\mathrm{4d}})}$
term in the high temperature expansion ($\tau\to 0$) of the corresponding Schur index
\cite{DiPietro:2014bca,Buican:2015ina,ArabiArdehali:2015ybk,
	ArabiArdehali:2023bpq}.\footnote{
    As an aside, we note that the vacuum character of $\mathcal{W}(4)$ satisfies a ninth order MLDE, whereas that of $\mathcal{A}(4)$ satisfies a third order twisted MLDE. The solution space of the former contains a solution of chiral weight $h=-5/16$. For $\mathcal{A}(4)$, this chiral weight appears among the solutions of the S-transformed, or conjugate, MLDE associated with the twisted MLDE.  This is not, however, the smallest chiral weight among the solutions of the corresponding differential equation. Such a phenomenon is typical for theories with
$c_{\mathrm{4d}}-a_{\mathrm{4d}}<0$, for which the solution governing the high-temperature asymptotics need not be the solution with minimum chiral weight
\cite{ArabiArdehali:2023bpq,Chen:2026vyz}.
    } This is a rather nontrivial consistency check.
    The four-dimensional central charges of the putative  SCFT were obtained from  Coulomb branch geometry considerations.
    While we used $c_{4d}$ as a crucial input in guessing the associated VOA,
    the fact that $a_{4d}$ is  also correctly reproduced (by both candidates) was certainly not foreordained.

Finally, let us consider the nilpotency index of both candidate VOAs in light of our general conjectures.
For the triplet algebra,
 we haven’t found a null of the requisite form \eqref{eq:nullvector} up to level $8$, implying that $\ioni (\mathcal{W}(4) ) > 4$.  For the doublet algebra, we have been able to determine $\ioni (\mathcal{A}(4) ) = 3$, as we now describe.
 The $\mathcal{A}(4)$ VOA is strongly generated by $T$, $L$ and $\tilde{L}$, where $L$ and $\tilde{L}$ are operators of dimension $5/2$. The $L\tilde{L}$ OPE is given as
\begin{equation}
	L(z)\tilde{L}(w)\sim\frac{1}{(z-w)^5}+\frac{-\frac{2}{5} T}{(z-w)^3}+\frac{-\frac{1}{5} T'}{(z-w)^2}+\frac{-\frac{1}{10} T''+\frac{2}{15} T^2}{z-w}\;.
\end{equation}
The remaining non-trivial OPEs are $TT$, $TL$ and $T\tilde{L}$ which admit the standard form. The algebra closes only for central charge $c=-\frac{25}{2}$. 
 There is a null vector of the form
\begin{equation}
	T^3+\frac{15 L'\tilde{L}}{2}-\frac{15 L \tilde{L}'}{4}-\frac{3 T T''}{4}-\frac{T^{(4)}}{16}
\end{equation}
and the vacuum character solves a third-order twisted MLDE. The first few terms for the vacuum character for $\mathcal{A}(4)$ are
\begin{equation}
	\chi_{\mathcal{A}(4)}= 1+q^2+2 q^{5/2}+q^3+2 q^{7/2}+2 q^4+2 q^{9/2}+2 q^5+4 q^{11/2}+4 q^6+\dots\;.
\end{equation}
The plethystic logarithm gives a better idea about generators and relations
\begin{equation}
	\text{PLog}(\chi_{\mathcal{A}(4)})=\frac{q^2+2 q^{5/2}-2 q^{9/2}-3 q^5-q^6+\dots}{1-q}\;.
\end{equation}
We can see that there is a $-q^6$ which is consistent with the fact that we found 
$T^3=0$ at that level.

According to the rank conjecture (\ref{conj:1}), the nilpotency index of the putative rank-one SCFT should be greater than or equal to 2, which is true for both candidate VOAs. On the other hand, as we have mentioned,  the  $IV^*_{Q=1}$ SCFT
should be reachable by RG flow from the ${\cal N}=3$ SCFT
with $\mathbb{C}^3/\mathbb{Z}_4$ moduli space,
whose nilpotency index is 4 (see Section \ref{sec:n3theories}). We conclude that the
doublet VOA
$\mathcal{A}(4)$ would be compatible with our RG flow conjecture (\ref{conj:2}),
while the triplet VOA $\mathcal{W}(4)$ would violate it. We take this as some evidence in favor of $\mathcal{A}(4)$.

%% file: sections/S_complexity.tex
\section{Theories with $\mathfrak{n}=1$ and $\mathfrak{n}=2$}
\label{sec:measureofcomplexity}

If our conjecture~1 is correct,
the nilpotency index provides a  ``measure of complexity'' of 4d ${\cal N}=2$ SCFTs which is more refined than their rank. This suggests the natural program of classifying SCFTs by increasing $\mathfrak{n}$. In this section we illustrate this idea in the simplest cases of $\mathfrak{n}=1$ and $\mathfrak{n}=2$.
For $\mathfrak{n}=1$ it is immediate to see that the only possibility is a collection of free hypermultiplets. For $\mathfrak{n}=2$ we will achieve a full classification under a certain natural assumption about the $R$-filtration of 
composite operators that may appear in the nilpotency null. We have explicitly checked that this assumption holds in all rank-one theories. Our calculations are significantly more straightforward than those needed to complete the full classification of rank-one theories by CB geometry methods -- this is natural as we find (under our $R$-filtration hypothesis) that theories with $\mathfrak{n}=2$ comprise the simplest subclass of rank-one theories, namely the Deligne series  and $SU(2)$ ${\cal N}=4$ SYM. The complexity of the requisite calculations will however rapidly increase with higher $\mathfrak{n}$.

Our analysis starts by  listing the operators that may appear in the $\phi\in C_2(\mathcal{V})$ part of the  nilpotency null
\begin{equation}
    T^\ioni+\phi,\;\;~\phi\in C_2(\mathcal{V})~.
\end{equation}
 
\begin{table}[]
    \centering
    \begin{tabular}{|c|c|c|}
    \hline
        Multiplet  & $h$ & $r$\\
    \hline
    \hline
       $\hat{\mathcal{B}}_R$  & $R$ & $0$ \\ 
       \hline
       $\mathcal{D}_{R(0,j_2)}$  & $R+j_2+1$ & $j_2+\frac{1}{2}$ \\ 
       \hline
       $\bar{\mathcal{D}}_{R(j_1,0)}$  & $R+j_1+1$ & $-j_1-\frac{1}{2}$ \\ 
       \hline
       $\hat{\mathcal{C}}_{R(j_1,j_2)}$  & $R+j_1+j_2+2$ & $j_2-j_1$ \\ 
       \hline
    \end{tabular}
    \caption{The table lists the 4d $\mathcal{N}=2$ superconformal multiplets containing Schur operators. Each multiplet contains one Schur operator, for which we indicate the quantum numbers.
     }
    \label{tab:schurmultiplets}
\end{table}
The operator  $\phi$ is a composite
of at least two strong generators with at least one derivative, or possibly a total derivative of a single strong generator (which may be needed to make $  T^\ioni+\phi$ a quasiprimary). For $\ioni=1$ and  $\ioni=2$, the chiral dimensions of the strong generators
that may appear in $\phi$ are restricted to $\frac{1}{2}$, $1$, $\frac{3}{2}$ and $2$. We can be precise about their possible 4d origin. 
Table~\ref{tab:schurmultiplets} displays how Schur operators fit in superconformal multiplets (in the notations of \cite{Dolan:2002zh}), giving their chiral dimensions and R-charges.
The relevant multiplets containing the Schur operators with chiral dimension $\frac{1}{2}$, $1$, $\frac{3}{2}$ and $2$ are:
\begin{equation}
\label{eq:schurneeded}
	\begin{split}
		h=\frac{1}{2}&\;:\;\textcolor{blue}{\hat{\mathcal{B}}_{\frac{1}{2}}}~,\\
		h=1 &\; :\; \hat{\mathcal{B}}_{1},\;  \textcolor{blue}{\mathcal{D}_{0(0,0)}\oplus \bar{\mathcal{D}}_{0(0,0)}}~,\\
		h=\frac{3}{2} &\; :\; \hat{\mathcal{B}}_{\frac{3}{2}},\;\; \textcolor{Green}{\mathcal{D}_{\frac{1}{2}(0,0)}\oplus \bar{\mathcal{D}}_{\frac{1}{2}(0,0)}}~,\;\; \xcancel{\textcolor{blue}{\mathcal{D}_{0(0,\frac{1}{2})}\oplus \bar{\mathcal{D}}_{0(\frac{1}{2},0)}}}~\\
		h=2 &\; :\; \hat{\mathcal{B}}_{2},\;\;\hat{\mathcal{C}}_{0(0,0)},~ \dots
	\end{split}
\end{equation} 
Multiplets that contain free fields are colored in blue: 
$\hat{\mathcal{B}}_{\frac{1}{2}}$ are the free hypermultiplets, 
$\mathcal{D}_{0(0,0)}\oplus\bar{\mathcal{D}}_{0(0,0)}$  the free vector multiplets,
while 
$\mathcal{D}_{0(0,\frac{1}{2})}\oplus\bar{\mathcal{D}}_{0(\frac{1}{2},0)}$ are multiplets containing exotic higher-spin free fields -- we have crossed them out as they are absent in a local SCFT.
The  multiplets in green, namely 
$\mathcal{D}_{\frac{1}{2}(0,0)}\oplus\bar{\mathcal{D}}_{\frac{1}{2}(0,0)}$, contain additional supercurrents, signalling a SUSY enhancement
to either $\mathcal{N}=3$ or $\mathcal{N}=4$ ($\mathcal{N} >4$ is not possible in a local SCFT).
The 
 $\hat{\mathcal{B}}_{1}$ multiplets contain flavor currents; the corresponding Schur operators map to affine Kac-Moody currents in the VOA.
 For $\mathfrak{n}=2$, an operator  $O$ with $h=2$
 may appear in $\phi$ as the total derivative $\partial^2  O$.
 Since it can appear only linearly in $\phi$, it must have zero $U(1)_r$ charge. This implies that it is a Higgs multiplet $\hat{\mathcal{B}}_{2}$ (since we are assuming that it is not the unique stress tensor multiplet $\hat{\mathcal{C}}_{0(0,0)}$). The $\dots$
 in \eqref{eq:schurneeded} indicate operators with non-zero $U(1)_r$ charge which do not play a role in our discussion.

We introduce the following naming conventions for the VOA generators corresponding to the Schur multiplets listed above (we have dropped various indices to lighten the notation):
\begin{equation}
\label{eq:multipletnotation}
	\begin{split}
            \xi&\sim \hat{\mathcal{B}}_{\frac{1}{2}}~,\\  
		J&\sim \hat{\mathcal{B}}_{1}~,~ \lambda\sim\mathcal{D}_{0(0,0)}\oplus\bar{\mathcal{D}}_{0(0,0)}\\
		W&\sim\hat{\mathcal{B}}_{\frac{3}{2}}~,~G\sim\mathcal{D}_{\frac{1}{2}(0,0)}\oplus \bar{\mathcal{D}}_{\frac{1}{2}(0,0)} ~\\
        O&\sim \hat{\mathcal{B}}_2~,~ T\sim\hat{\mathcal{C}}_{0(0,0)}
	\end{split}
\end{equation}
The operators $\xi$, $J$, $W$, $O$, $T$ are bosonic while $\lambda$, $G$ are fermionic. As VOA operators, $\xi$ are symplectic bosons, $J$ are AKM currents, $\lambda$ are symplectic fermions, $G$ are dimension $\frac{3}{2}$ supercurrents, $W$ is a chiral operator with dimension $\frac{3}{2}$ and $T$ is the stress tensor. Finally $O$ is a dimension $2$ operator corresponding to the $\hat{\mathcal{B}}_2$ multiplet.

\subsection{$\mathfrak{n}=1$}
The case of $\mathfrak{n}=1$ is straightforward. Based on chiral dimensions, the only multiplets that one can consider are $\hat{\mathcal{B}}_{\frac{1}{2}}$ and $\hat{\mathcal{B}}_{1}$. The general form of the null is
\begin{equation}	T+\sum_{i,j}\Omega_{ij}\xi^i\partial \xi^j ~,
\end{equation}
where $\Omega^{ij}$ is the symplectic form corresponding to the OPE $\xi^i(z)\xi^j(w)\sim\frac{\Omega^{ij}}{z-w}$. $\partial J$ is disallowed because it is not a quasiprimary. We  conclude at once that the only theories with $\ioni=1$ are collections of free hypermultiplets. 
\subsection{$\ioni=2$}
As have discussed in section \ref{sec:sympferm},
the free vector multiplet has $\ioni=2$ and is the simplest example for a theory with $\ioni=2$.
Let us then consider theories with no free decoupled sector. The schematic form of the nilpotency null  is 
\begin{equation} \label{schematicN}
	\mathfrak{N}=\left(T^2+\partial^2 T\right)+\left(\partial J\partial J+\partial^2J J\right)+ \partial G G+ \partial W W+\partial^2O~ ,
\end{equation}
where we have set to one the non-trivial coefficients in front of each term  and it is understood that flavor indices are contracted\footnote{Notice that since $W$ (respectively $G$) has dimension $3/2$ and is bosonic (fermionic) it must transform in a pseudoreal (real). This guarantees that quasi-primaries which are flavor singlet of the schematic form $\partial G G+ \dots$ and $\partial W W+\dots$  do exist.} and the expression is a sum of quasi-primaries\footnote{The strange looking term $\partial^2O$ has the only purpose of completing the remaining terms to quasi-primaries.}.
Like all nulls, $\mathfrak{N}$ should be a Virasoro primary. Its  OPE with $T$ takes the  form
\begin{equation}
\label{eq:OPETnull}
    T(z)\mathfrak{N}(w)\sim \frac{(c+\frac{22}{5})T(w)+\# J(w)^2+ O(w)}{(z-w)^4}+\frac{4\mathfrak{N}(w)}{(z-w)^2}+\frac{\partial\mathfrak{N}(w)}{(z-w)}\,.
\end{equation}
Here we have assumed that  $\mathfrak{N}$ is a quasiprimary (which is always easy to arrange) so that the third order pole vanishes.
For $\mathfrak{N}$ to be a Virasoro primary,
 the forth order pole  must also vanish, i.e.~the residue must be a null state. We observe that $O$ and
 $(c+\frac{22}{5})T(w)+\# J(w)^2$ must be separately null, as $O$ is by assumption a strong generator with $R=2$ while $T$ has $R=1$.
We conclude that there is no $\partial^2 O$  in (\ref{schematicN}) to begin with, and that for $c \neq -22/5$ the stress tensor must be the Sugawara stress tensor for the AKM algebra. 

When $c=-\frac{22}{5}$, the bound of \cite{Lemos:2015orc} excludes a nontrivial flavor symmetry \footnote{For $c_{4d}=-12c_{2d}=11/30$, the bound in equation (2.20) of \cite{Lemos:2015orc} reduces to $3k_{\text{4d}}\dim_G\leq 0$, where $\dim_G$ denotes the dimension of the flavor symmetry $G$. Since $k_{\text{4d}}>0$ for a nontrivial flavor symmetry, the inequality implies $\dim_G=0$. }. Therefore, for this value of central charge, the 4d theory cannot have a flavor symmetry and the fourth order pole is automatically zero. 

We now  split the discussion by first considering the cases with extended supersymmetry ($\mathcal{N}=3$ and 
$\mathcal{N}=4$) and then the pure $\mathcal{N}=2$
cases.

\subsubsection{$\mathcal{N}=3$  and $\mathcal{N}=4$}
\label{sec:n34susy}
In the following we will show by a simple argument that the only theory with extended supersymmetry and $\n =2$ is $SU(2)$ $\mathcal{N}=4$ SYM.
Extra supersymmetries correspond to $\mathcal{N}=2$ supermultiplets of type $\mathcal{D}_{\frac{1}{2}(0,0)}\oplus \bar{\mathcal{D}}_{\frac{1}{2}(0,0)}$. In a local SCFT,  supersymmetry can be enhanced to $\mathcal{N}=3$ or $\mathcal{N}=4$. These theories do not have global flavor symmetries
(i.e.~symmetries commuting with their superconformal algebras),
see \cite{Aharony:2015oyb,Cordova:2016emh}.
The $\mathcal{N}=3$ and $\mathcal{N}=4$ stress-tensor supermultiplets in 4d give rise, under the chiral algebra map, to the strong generators of the $\mathcal{N}=2$ superconformal algebra (SCA) and the small $\mathcal{N}=4$ SCA, see \cite{Bonetti_2019} for more details.
The $\mathcal{N}=2$ SCA has generators
\begin{equation}
\mathcal{T},\;\;\mathcal{J},\;\;\mathcal{G},\;\;\tilde{\mathcal{G}}~,
\end{equation}
where $\mathcal{J}$ is associated to a $U(1)$ flavor symmetry.
The  small $\mathcal{N}=4$ SCA has  generators 
\begin{equation}
\begin{split}
    \text{Stress Tensor} &\;\;\;T~,  \\ 
    \text{AKM currents} &\;\;\; J_{ab}\;\;\;\;\;\;a,b=1,2~,\\ 
    \text{Supercurrents} &\;\;\; G_a, \tilde{G}_a\;\;\;a=1,2~,
\end{split}
\end{equation}
where $J_{ab}=J_{ba}$ are $SU(2)$ current.
The closure of the OPE of the small $\mathcal{N}=4$ super-Virasoro algebra implies that the AKM level $k$ is related to the central charge $c$ as $c=6k$. This will play an important role momentarily.
 
The schematic form of the null is given by
\begin{equation}
\label{nullhere}
	\left(T^2+\partial^2 T\right)+\left(\partial J\partial J+\partial^2J J\right)+ \partial G G+\partial W W~,
\end{equation}
where $J$ is a $U(1)$ or $SU(2)$ current generator depending on whether the theory has $\mathcal{N}=3$ or $\mathcal{N}=4$ SUSY and, at this stage, we include the possibility of having extra generators denoted by $W$ as above.
As discussed in the previous section, a necessary condition to have nilpotency index two is that the stress tensor coincides with the Sugawara stress tensor.

For $\mathcal{N}=3$, imposing the Sugawara condition means that\footnote{Adding a term of the form $\partial \mathcal{J}$ would spoil the OPE of $T$ with the supercurrents.} $T\propto \mathcal{J}^2$. This implies that the central charge $c=1$. Because $c_{2d}=-12c_{4d}<0$, such a solution is not allowed and therefore for all $\mathcal{N}=3$ theories, $\ioni\neq 2$. For the $\mathcal{N}=4$ theories, the Sugawara condition gives $c=\frac{3k}{k+2}$, which together with the relation $c=6k$ mentioned above implies (discarding the trivial solution $c=0$) that $c=-9$ (and $k=-\frac{3}{2}$). The final step to show that the (simple quotient) of the small $\mathcal{N}=4$ super-Virasoro algebra at $c=-9$ is the only VOA associated to theories with extended supersymmetry in 4d and $\ioni=2$ is the following.
As discussed above we must set to zero the combination $T-T_{\text{Sug}}$. This means that we are taking the simple quotient of the small $\mathcal{N}=4$ super-Virasoro algebra at $c=-9$.
This VOA admits only one module other than the vacuum module, see \cite{Adamovic:2014lra}. This extra module has weight $h=-\tfrac{1}{2}$, and cannot be associated a generator of the VOA originating from a four dimensional multiplet. 
This implies that  the small $\mathcal{N}=4$ SCA at $c=-9$ does not admit any further extension (so that there can be no $W$s in \eqref{nullhere}) and therefore the only theory with extended supersymmetry and $\ioni=2$ is the $\mathcal{N}=4$ $SU(2)$ SYM.

\subsubsection{$\mathcal{N}=2$}
\label{sec:neq2}
Now we come back to the discussion of pure $\mathcal{N}=2$ theories. We will assume that the composite $\partial W W$ has $R=3$,
which is its ``natural'' value  -- 
 the filtration  of the composite doesn't drop. As we discuss below, we have verified this assumption in all rank-one theories. The same assumption can used to establish lower bounds on $\ioni$, see table~\ref{tab:ionirankone}. 
This  eliminates operators of the form $\partial W W$ in the nilpotency null, because the remaining operators have $R$-assignment  at most $2$. We thus have the schematic ansatz
\begin{equation}
\label{eq:schemenull}
\mathfrak{N}=	\left(T^2+\partial^2 T\right)+\left(\partial J\partial J+\partial^2J J\right).
\end{equation}
We now use the crucial fact (which we proved above) that
$T$ is a Sugawara stress tensor. A first consequence is that the Lie algebra must consist of a single simple factor. Indeed,  consider a theory for which the affine Lie algebra is a direct sum $\mathfrak{g}_1\oplus\mathfrak{g}_2$. We denote the respective currents generators as $J_1$ and $J_2$. Because the stress tensor is Sugawara, the stress tensor $T$ is a sum of Sugawara stress tensors of the individual AKMs,
\begin{equation}
	T=T_1+T_2,\;\;\;T_i\propto(J_i)^2\,.
\end{equation}
Clearly
\begin{equation}
	T^2=T_1^2+T_2^2+2T_1T_2\;.
\end{equation}
Since there are no non-trivial OPEs between the currents $J_1$ and $J_2$, any null can be decomposed in terms of the nulls of the individual AKMs $\mathfrak{g}_1$ and $\mathfrak{g}_2$. Therefore 
the mixed term $T_1T_2$ cannot belong to $C_2(\mathcal{V})$. Therefore we can restrict to simple Lie algebras. The expression \eqref{eq:schemenull} has zero norm if the coefficients are chosen as below,
\begin{equation}
	T^2-\frac{3}{10}\partial^2 T+\frac{3\kappa_{ab}}{5(h^\vee+6k)}\left(3\partial J^a \partial J^b-2\partial^2 J^a J^b\right)\;.
\end{equation}
The condition that the norm is zero also gives an expression for the central charge $c$ in terms of Lie-algebraic data,
\begin{equation}
	\label{eq:nullconstraint}	
	c=-\frac{11}{5}\left(1+\sqrt{1+\frac{180 \,\text{dim}(\mathfrak{g}) (-2 k)}{121 (3 (-2 k )-h^{\vee})}}\right)\;,
\end{equation}
where $h^{\vee}$ is the dual Coxeter number and dim$(\mathfrak{g})$ is the dimension of the Lie algebra.
 Equation \eqref{eq:nullconstraint} is same as the saturation of the bound in equation 4.9 of \cite{Beem_2018},
\begin{equation}
	-\frac{c}{12}\leq\frac{11}{60}\left(1+\sqrt{1+\frac{180 \,\text{dim}(\mathfrak{g}) (-2 k)}{121 (3 (-2 k )-h^{\vee})}}\right)\;.
\end{equation}
Solving \eqref{eq:nullconstraint} and the Sugawara condition $c=\frac{k\, \text{dim}(\mathfrak{g})}{k+h^\vee}$ gives the expression of the central charges $c$ and $k$ in terms of the dimension and dual-Coxeter number  of the Lie-algebra
\begin{equation}
	\label{eq:ccentral}
	c=\frac{1}{10} \left(-22-\text{dim}(\mathfrak{g})-\sqrt{(\text{dim}(\mathfrak{g})+2) (\text{dim}(\mathfrak{g})+242)}\right)\;,
\end{equation}
\begin{equation}
	\label{eq:kcentral}
	k=-\frac{ h^\vee\left(\text{dim}(\mathfrak{g})+\sqrt{(\text{dim}(\mathfrak{g})+2) (\text{dim}(\mathfrak{g})+242)}+2\right)}{12 (\text{dim}(\mathfrak{g})+2)}\;.
\end{equation}
With this information, we simply go over the list of simple Lie algebras and find the possible solutions to equations \eqref{eq:ccentral} and \eqref{eq:kcentral} with the requirement that the central charges are rational (see \cite{Rastelli:2023sfk}). This gives a list of theories which includes Deligne rank-one theories. Further imposing four dimensional unitarity bounds on $k$ (see \cite{Beem_2015, Beem_2018,Beem:2018duj}), restricts us precisely to the Deligne rank-one SCFTs.  In summary, using our $R$-filtration hypothesis, the only possible theories $\ioni=2$ are the Deligne rank-one theories and $\mathcal{N}=4$ $SU(2)$ SYM.

\subsubsection{Rank-one theories}
\label{sec:rankonespecialize}
We have managed to compute $\n$ for some rank-one theories; for others we were able to
put a lower bound.
 In
Table \ref{tab:ionirankone} we  label rank-one theories by their Coulomb branch singularity and flavor symmetry (see \cite{Argyres:2015ffa, Argyres:2015gha,Argyres:2016xmc,Argyres:2016xua,Argyres:2020nrr} for more details). We have omitted theories which have a trivial Higgs branch, which include the free vector mutliplet and a few exotic cases (one such exotic case is discussed in Section \ref{sec:IVrankone}). The theories are arranged in blocks, where any theory in a given block can be reached by an RG flow from theories above it, in the same block. For Deligne rank-one theories, the analysis was already done in Sections \ref{sec:drankone} and \ref{sec:neq2}, where $\ioni$ was shown to be $2$.

If $\ioni=2$, the stress tensor is necessarily Sugawara. Only Deligne theories, $[III^*,C_3 A_1]$, $[IV^*,C_2 U_1]$ and $\textcolor{blue}{[I_0^*,C_1]}$ ($SU(2)$ $\mathcal{N}=4$ SYM) have their stress tensors as Sugawara and therefore we deduce that $\ioni$ is greater than $2$ for the remaining theories.

We can put a bound on $\ioni$ for $[III^*,C_3 A_1]$ and $[IV^*,C_2 U_1]$ by explicitly checking our $R$-filtration hypothesis. In Section \ref{sec:neq2}, we assumed that the R-charge is $3$ for terms of type $\partial WW$. For all the rank-one theories, we explictly checked this using the free field realization \cite{Beem:2024fom} and it is indeed true. All theories, except the theories in the last two blocks have a dimension $\frac{3}{2}$ generator corresponding to a $\hat{\mathcal{B}}_{\frac{3}{2}}$ multiplet. For $[II^*,A_2]$, $[III^*,U_1]$ and $[II^*,U_1]$, there are no strong generators corresponding to the $\hat{\mathcal{B}}_{\frac{3}{2}}$ multiplets, so a nilpotency null would have to take the form \eqref{eq:schemenull}; since the stress tensor of these theories is not Sugawara, $\ioni>2$. Such a multiplet is present for the $[II^*,C_5]$, $[III^*,C_3 A_1]$, $[IV^*,C_2 U_1]$, $[II^*,A_3]$, $[III^*,A_1U_1]$ and $[IV^*,U_1]$. But the singlet of the form $\partial W W$ does not drop in filtration and therefore $\ioni>2$ for all these theories, which include $[III^*,C_3 A_1]$ and $[IV^*,C_2 U_1]$ as well. 

\begin{table}[t]
	\begin{adjustbox}{center}
		\begin{tabular}{ |c|c| } 
			\hline
			Theory & Nilpotency index $\ioni$ \\
			\hline
			\hline
			Deligne rank-one & $\ioni=2$ \\
			\hline
			\hline
			$[II^*,C_5]$ & $\ioni >2$	\\
			$[III^*,C_3 A_1]$ & $\ioni>2$	\\
			$[IV^*,C_2U_1]$ & $\ioni>2$	\\
			\textcolor{blue}{$[I_0^*,C_1]$} & $\ioni=2$\\
			\hline
			\hline
			$[II^*,A_3]$ & $\ioni >2$	\\
			$[III^*,A_1 U_1]$ & $\ioni>2$	\\
			\textcolor{green}{$[IV^*,U_1]$}  & $\ioni=3$\\
			\hline
			\hline
			$[II^*,A_2]$  & $\ioni>2$\\
			\textcolor{green}{$[III^*,U_1]$}  & $\ioni=4$\\
			\hline
        \hline
			\textcolor{green}{$[II^*,U_1]$}  & $\ioni>2$\\
			\hline
		\end{tabular}
		
	\end{adjustbox}
	
	\caption{The nilpotency index for rank-one theories. The theories are organized into five blocks. Within each block, a theory can be reached by a renormalization group flow triggered by relevant deformations from theories above it. We know that $\ioni=2$ for   all the Deligne rank-one theories. For the remaining blocks, we denote the theory by its flavor symmetry and Coulomb branch singularity. The blue theory is the $SU(2)$ $\mathcal{N}=4$ SYM and the green theories have $\mathcal{N}=3$ SUSY.}
\label{tab:ionirankone}
\end{table}

%% file: sections/S_RG.tex
\section{Renormalization group flows}
\label{sec:RGFlow}
We now discuss the behavior  of the nilpotency index $\n$ under RG flows. The upshot is that we have collected some evidence
for conjecture \ref{conj:2}, that $\n$ is non-increasing as one flows from a UV to an IR SCFT.

We consider two classes of RG flows.
In the first class, the RG flow is triggered by a relevant deformation. In order to land on a nontrivial IR SCFT, the relevant deformation must sometimes be accompanied by moving to a certain fine tuned point on the Coulomb branch.
In the second class, the flow is triggered by moving on the Higgs branch. In this case the theory in the IR is generally an interacting theory along with decoupled free hyper or vector multiplets. Since Higgsing in 4d theory is implemented as a DS-reduction of the VOA, a special case of conjecture~\ref{conj:2} is conjecture~\ref{conj:3}: $\ioni$ is non-increasing under DS reduction.

As already mentioned in the introduction, conjecture~2 automatically implies conjecture \ref{conj:1}. Indeed, at generic point on the Coulomb branch, the theory consists of just $\rank$ free vector multiplets, whose nilpotency index equals $\rank+1$. If $\ioni$ is non-increasing, we must have $\ioni\geq\rank+1$.

The IR theory may consist of a tensor product of decoupled  SCFTs. In those cases, one should compute $\ioni$ using the formula described in Section \ref{sec:tenstheor}. Let us consider a case where the UV theory $\mathcal{T}_{\text{UV}}$ has nilpotency index $\ioniUV$ and in the infrared we have a theory $\mathcal{T}_{\text{IR}}$ consisting of an interacting theory $\mathcal{T}_{\text{int}}$, $n_h$ free hypermultiplets and $n_v$ free vector multiplets. Let us call the nilpotency index in the infrared as $\ioniIR$. Using equation \eqref{eq:ntensor}  for tensor of theories, we obtain $\ioniIR=\ioni_{\text{int}}+n_v$ (note that the trivial theory has $\ioni=1$.). Therefore we can see that the presence of hypermultiplets as decoupled sectors doesn't have any effect on the total $\ioni$, on the other hand presence of each free vector multiplet increases $\ioniIR$ by one. 

Before moving on to examples, we provide a sketch of a scheme theoretic proof of conjecture 2 in the special case of nilpotent Higgsings\footnote{We would like to thank Tomoyuki Arakawa for suggesting this.}. 

\subsection{Sketch of a scheme theoretic proof}
\label{sec:schemeproof}

In this section, we sketch a proof of Conjecture 2 for nilpotent
Higgsings by working at the level of schemes. We first recall that, for a
VOA $\mathcal V$, the associated scheme is defined as
\begin{equation}
   \widetilde{\mathcal{X}}_{\mathcal V}
   :=
   \operatorname{Spec}(\mathcal R_{\mathcal V}) ,
\end{equation}
where \(\mathcal R_{\mathcal V}=\mathcal V/C_2(\mathcal V)\), see \eqref{ZhuC2}. Its reduced scheme is the usual
associated variety,
\begin{equation}
   \mathcal{X}_{\mathcal V}
   =
   \left(\widetilde{\mathcal{X}}_{\mathcal V}\right)_{\mathrm{red}} .
\end{equation}
For our purposes it is crucial to work with the associated scheme
\(\widetilde{\mathcal{X}}_{\mathcal V}\), rather than with the associated
variety \(\mathcal{X}_{\mathcal V}\), since nilpotency properties in
\(\mathcal R_{\mathcal V}\) are lost after passing to the reduced scheme.

The main input for the proof is Theorem 9.11 of \cite{arakawa2021arc}, which makes
precise the compatibility between performing a Drinfeld--Sokolov reduction and passing to 
associated schemes. More explicitly, let $\mathcal V$ contain a sub VOA of affine $\mathfrak g$ currents\footnote{
More precisely $\mathcal{V}$ is a
$V^k(\mathfrak g)$-vertex algebra on which $\mathfrak g[[t]]$ acts locally finitely.} and let $\operatorname{DS}_f(\mathcal V)$ denote its
Drinfeld--Sokolov reduction with respect to a nilpotent element $f$.
If \(S_f\subset \mathfrak g^*\) is the corresponding Slodowy slice\footnote{
	Let $f$ be a nilpotent element of $\mathfrak{g}$, $\{e,h,f\}$ an $\mathfrak{sl}_2$-triple in $\mathfrak{g}$ associated with $f$. Then the \textit{Slodowy slice} at $f$ is given as
	\begin{equation}
		\mathcal{S}_f=f+\mathfrak{g}^e\subset\mathfrak{g}\simeq \mathfrak{g}^*~,
	\end{equation}
	where $\mathfrak{g}^e$ is the centralizer of $e$ in $\mathfrak{g}$.
}, then \cite{gan2002quantization,arakawa2021arc}
\begin{equation}
   \widetilde {\mathcal{X}}_{\operatorname{DS}_f(\mathcal V)}
   \simeq
   \widetilde{\mathcal{X}}_{\mathcal V}\times_{\mathfrak g^*} S_f .
\end{equation}
See \cite{arakawa2015associated} for the corresponding statement of DS-reduction for associated variety.
Equivalently, at the level of $C_2$-algebras
\begin{equation}
\label{theoreminALGEBRA}
  \mathcal{R}_{\text{IR}}= \mathcal R_{\operatorname{DS}_f(\mathcal V)}
   \simeq
   \mathcal R_{\mathcal V}
   \otimes_{\mathbb C[\mathfrak g^*]}
   \mathbb C[S_f] .
\end{equation}
To digest this formula, lets see it at work in the well studied example 
$\mathcal{V}_n:=\mathcal{V}_{-\frac{4n}{2n+1}}(\mathfrak{sl}_2)$. The variety associated to this family of VOAs is $\mathbb{C}^2/\mathbb{Z}_2$ for all values of $n=1,2,\dots$ but the corresponding $C_2$-algebras are different \cite{Beem_2018}
\begin{equation}
\mathcal R_{
\mathcal{V}_{n}
}
   \simeq
   \frac{\mathbb C[e,h,f]}
   {\bigl\langle e\,Q^n,\; h\,Q^n,\; f\,Q^n \bigr\rangle}\,,
   \qquad
   Q=h^2+4ef ,
\end{equation}
where $e,f,h$ denote the standard generators of $\mathfrak{sl}(2)$.
In particular $Q$, which is associated to the stress tensor, is nilpotent with $Q^{n+1}=0$ (and $Q^n\neq 0$).
Intersecting with the Slodowy slice has the effect of setting $e=u$, $h=0$ and $f=1$. This implies that  in this case $\mathcal{R}_{\text{IR}}$ is generated by $u$. Moreover the three relations in $\mathcal{R}_{\mathcal{V}_n}$ reduce to the single condition $u^n=0$. This procedure correctly reproduces the $C_2$-algebra $\mathbb{C}[u]/\langle u^n\rangle$ of the $C_2$-cofinite $(2,2n+1)$ Virasoro VOAs. In this example, the $C_2$-class of the UV stress tensor in $\mathcal{R}_{\mathcal{V}_n}$ is identified with the $C_2$-class of the IR stress tensor in $\mathcal{R}_{\text{IR}}$. 

If this identification holds in general, then
\eqref{theoreminALGEBRA} immediately implies that the nilpotency index of
the stress tensor cannot increase: restriction to the Slodowy slice can
impose additional relations, but cannot remove relations already
satisfied by the UV $C_2$-class.

The only additional point is thus to identify the class of the conformal vector under the
isomorphism \eqref{theoreminALGEBRA}. The standard conformal vector of
the BRST complex is of the form (see for example \cite{Beem:2014rza,deBoer:1993iz,deBoer:1992sy})
\begin{equation}
   T_{\mathrm{BRST}}
   =
   T_{\mathcal V}
   +
   T_{\mathrm{gh}}
   +
   \partial J,
\end{equation}
where $T_{\mathcal{V}}$ is the stress tensor of the UV theory, $T_{\text{gh}}$ is contribution from the ghosts and $J$ denotes the $\mathfrak{u}(1)$ generator in the $\mathfrak{sl}(2)$. The cohomology class of $T_{\text{BRST}}$ is the conformal vector of
\(\operatorname{DS}_f(\mathcal V)\) (see
\cite[Section~9.1.2]{arakawa2021arc}). Both the ghost contribution and
the improvement term have positive Li degree. Since
\(\mathcal R_{\mathcal V}=F^0\mathcal V/F^1\mathcal V\), they vanish
after passing to the \(C_2\)-algebra. Consequently, under the
identification \eqref{theoreminALGEBRA}
one has
\begin{equation}
   [T_{\operatorname{DS}_f(\mathcal V)}]
   =
   [T_{\mathcal V}]\otimes 1.
\end{equation}
Thanks to  this identification, any relation $[T_{\mathrm{UV}}]^n=0$ implies $[T_{\mathrm{IR}}]^n=0$, and therefore the nilpotency index of the stress tensor cannot increase under DS reduction.

\subsection{RG flows from relevant deformations}
Here we test our conjecture for RG flows where the IR theory is obtained by turning on a relevant deformation and then tuning the Coulomb branch vev to a singular point on the Coulomb branch. The arrows are directed from the UV theory to the IR theory and the subscript denotes the nilpotency index.

\paragraph{\underline{ Deligne rank-one theories}} 
These theories are very uniform and the index stays constant along the flow connecting them
\begin{equation}
	(\mathfrak{e}_8)_{\mathfrak{n}=2} \to (\mathfrak{e}_7)_{\mathfrak{n}=2} \to (\mathfrak{e}_6)_{\mathfrak{n}=2}\to(\mathfrak{d}_4)_{\mathfrak{n}=2}\to(\mathfrak{a}_2)_{\mathfrak{n}=2}\to (\mathfrak{a}_1)_{\mathfrak{n}=2}\to (\mathfrak{a}_0)_{\mathfrak{n}=2}\;.
\end{equation}

\paragraph{\underline{ Argyres-Douglas theories}}

Next, we discuss Argyres-Douglas theories $(A_{k-1},A_{N-1})$. These theories flow to $(A_{k-1},A_{N-3})$ in the IR after turning on a relevant deformation (see \cite{Xie_2013} for more details), 
\begin{equation}
	(A_{k-1},A_{N-1})\to (A_{k-1}, A_{N-3})\;.
\end{equation}
We will check special cases of the above flow. Theories of the form $(A_1,A_{2n})$ and $(A_1,A_{2n+1})$ belong to this class with $k=2$, $N=2n$ and $N=2n+1$ respectively. 
\begin{equation}
	(A_1,A_{2n})_{\ioni=n+1}\to (A_1,A_{2n-2})_{\ioni=n}\;,
\end{equation}
\begin{equation}
	(A_1,A_{2n+1})_{\ioni=n+1}\to (A_1,A_{2n-1})_{\ioni=n}\;.
\end{equation}
Let us apply it to $(A_2,A_3)\sim (A_1,E_6)$ and  $(A_2,A_4)\sim (A_1,E_8)$. The flows are as follows
\begin{equation}
	(A_2,A_3)_{\ioni=4}\to(A_2,A_1)_{\ioni=2}\sim (A_1,A_2)_{\ioni=2}\;,
\end{equation}
\begin{equation}
	(A_2,A_4)_{\ioni=5}\to(A_2,A_2)_{\ioni= 2}\sim (A_1, D_4)_{\ioni= 2}\;.
\end{equation}
Let us also consider the following two series for RG flows described in \cite{Xie_2013}
\begin{equation}
(A_1,E_6)_{\ioni=4}\to(A_1,D_4)_{\ioni=2}\to(A_1,A_2)_{\ioni=2}\to (II_{3,1})_{\ioni=1}\;,
\end{equation}
\begin{equation}
	(A_1,A_1)_{\ioni=1}\leftarrow (A_1,D_6)_{\ioni=3}\rightarrow (A_1,A_4)_{\ioni=3}\rightarrow (A_1,A_2)_{\ioni=2}\;.
\end{equation}
Both $II_{3,1}$ and $(A_1,A_1)$ denote the theory of one free full hypermultiplet.
\paragraph{\underline{rank-one $IV^*_{Q=1}$ theory}}
We have also used the conjecture to provide a candidate VOA for this theory. We have already discussed this is detail in Section \ref{sec:IVrankone}, so we do not repeat the discussion here.

\subsection{RG flows from Higgsing}

In this section, we consider RG flows where the IR theory is obtained by Higgsing the UV theory. Generically, Higgsing a theory leads to an interacting theory along with decoupled free hypermultiplets and free vector multiplets. In all the examples of Higgsing discussed below, we will have a decoupled set of hypermultiplets or vector multiplets along with an interacting theory.

\paragraph{\underline{Deligne rank-one}}
The simplest set of examples are the rank-one Deligne series. The Deligne theories have $\ioniUV=2$. In the infrared, we find just free hypers which have $\ioniIR=1$.

\paragraph{\underline{rank-one theories with $\mathcal{N}\geq 3$ SUSY}}
As discussed in Section \ref{sec:n3theories}, rank-one $\mathcal{N}\geq 3$ SCFTs are labelled by a positive integer $\ell$ and have moduli space $\mathbb{C}^3/\mathbb{Z}_\ell$. $\ioniUV=2,3$ and $4$ for $\ell=2,3$ and $4$ respectively. The theory obtained after Higgsing is just a theory of a single free hyper and a single free vector multiplet. Therefore $\ioniIR=2$ and $\ioniUV\geq \ioniIR$ for all the above cases.

\paragraph{\underline{Argyres-Douglas}}
Another case of Higgsing discussed in \cite{Beem_2019} is the RG flow from $(A_1,D_{2n+1})$ (whose HB is $\mathbb{C}^2/\mathbb{Z}_2$) to $(A_1,A_{2n-2})$ (whose HB is a point) and a free hyper. In this case we have $\ioniUV=n+1$ and $\ioniIR=n$. Similarly one can Higgs
 $(A_1,D_{2n+2})$ (whose HB has quaternionic dimension $2$) to $(A_1,A_{2n-1})$ (whose HB is $\mathbb{C}^2/\mathbb{Z}_n$). Also in this case the nilpotency index decreases by one unit: $\ioniUV=n+1$ and $\ioniIR=n$.

\paragraph{\underline{Class S}}
Although for this class of theories, we only have a conjectural value of $\ioni$ from the observations in the Macdonald index, we can still use this value to test conjecture \ref{conj:2}. Starting with a theory with $s$ punctures, the IR theory is obtained by closing a puncture, which is also Higgsing. It is  straightforward to see that $\ioniUV=s-2$ and $\ioniIR=s-3$. 

\paragraph{\underline{Deligne rank-two $\mathfrak{d}_4$}}
Similarly for Deligne rank-two $\mathfrak{d}_4$, we  can use the conjectural value of $\ioniUV=3$ from the observations in the Macdonald index. Higgsing the theory gives two copies of Deligne rank-one $\mathfrak{d}_4$ ($\ioni=2$) theories and free hypers. The combined IR system has $\ioniIR=3$ consistent with our conjecture.